\documentclass[iop]{emulateapj}
\usepackage{natbib}
\usepackage{longtable}
\usepackage{amsmath}
\usepackage{epsfig}
\usepackage{float}
\usepackage{textcomp}
\usepackage{color}
\usepackage{apjfonts}

\slugcomment{Received 2015 March 25; accepted 2015 September 17; published 2015 October 23}

\shorttitle{Young Stellar Objects in W49}
\shortauthors{Saral et al.}

\begin{document}

\title{Young Stellar Objects in the Massive Star-Forming Region W49}

\author{
G. Saral\altaffilmark{1,2},
J. L. Hora\altaffilmark{1},
S. E. Willis\altaffilmark{1},
X. P. Koenig\altaffilmark{3},
R. A. Gutermuth\altaffilmark{4},
A. T. Saygac\altaffilmark{5}
}

\altaffiltext{1}{Harvard-Smithsonian Center for Astrophysics, 60 Garden Street, Cambridge, MA 02138, USA}
\altaffiltext{2}{Istanbul University, Graduate School of Science and Engineering, Bozdogan Kemeri Cad. 8, Vezneciler-Istanbul-Turkey}
\altaffiltext{3}{Yale University, Department of Astronomy, 208101, New Haven, CT 06520-8101, USA}
\altaffiltext{4}{University of Massachusetts, Department of Astronomy, Amherst, MA 01003, USA}
\altaffiltext{5}{Istanbul University, Faculty of Science, Astronomy and Space Sciences Department, Istanbul-Turkey}

\begin{abstract}
We present the initial results of our investigation of the star-forming complex W49, one of the youngest and most luminous massive star-forming regions in our 
Galaxy. We used \textit{Spitzer}/Infrared Array Camera (IRAC) data to investigate massive star formation with the primary objective of locating a representative 
set of protostars and the clusters of young stars that are forming around them. We present our source catalog with the mosaics from the IRAC data. 
In this study we used a combination of IRAC, MIPS, Two Micron All Sky Survey, and UKIRT Deep Infrared Sky Survey (UKIDSS) data to identify and 
classify the young stellar objects (YSOs). We identified 232 Class 0/I YSOs, 907 Class II YSOs, and 74 transition disk candidate objects using color$-$color and 
color$-$magnitude diagrams. In addition, to understand the evolution of star formation in W49, we analyzed the distribution of YSOs in the region to identify 
clusters using a minimal spanning tree method. The fraction of YSOs that belong to clusters with $\geq$$7$ members is found to be $52\%$ for a cutoff 
distance of 96$\arcsec$, and the ratio of Class II/I objects is 2.1. We compared the W49 region to the G305 and G333 star-forming regions and concluded that the 
W49 has the richest population, with seven subclusters of YSOs. 
\end{abstract}


\section{Introduction}

It is generally accepted that approximately $70-90\%$ of stars form in groups and clusters embedded in collapsing molecular clouds \citep{lal03,brs10}. 
Observations of nearby star-forming regions have revealed how a low-mass star can form in an isolated environment. 
However, massive stars play a vital role in the star formation process in embedded clusters, yet their own formation and their effects on subsequent 
generations of star formation are not well understood. Massive stars form relatively quickly compared to low-mass stars and reach the main sequence still 
embedded in their natal clump. The short duration of this phase means that these objects are relatively rare and thus typically more distant, making them 
difficult to study. In addition, the embedded clusters hosting those massive stars are hidden inside their natal molecular cloud and can only be observed 
at infrared (IR) and millimeter wavelengths. 

The launch of the \textit{\textit{Spitzer Space Telescope}} in 2004 \citep{wer04} has had a big impact on our understanding of the star formation process. 
Many nearby star-forming regions (e.g. Taurus complex, M16, M17, NGC 6334) have been studied \citep{tor09,kuh13,wil13}, and many methods have been suggested 
and used to classify the young stellar objects (YSOs) through their spectral indices \citep{lad87,rob06} or color$-$color diagrams \citep{gut08,gut09}. The combination of 
\textit{Spitzer}-Infrared Array Camera (IRAC) \citep{all04,faz04,gut08} and near-IR data has been a powerful tool to identify and classify YSOs.

We performed a detailed investigation of the \object{W49} star-forming region, which is located within one of the most massive giant molecular clouds (GMCs) 
\citep[$M_\mathrm{gas}\sim10^6$ M$_\odot$;][]{sim01} in the Galaxy and hosts many massive protostars and clusters of young stars that are forming around them. 
Since it is not possible to observe a single YSO at various stages of its evolution, we have to observe large numbers of YSOs to study the pre-main-sequence 
evolution statistically \citep{lad92,all04,gut09}. 

The \object{W49} star-forming region was discovered as a radio continuum source by \citet{wes58} and lies in the Galactic plane ($l, b = 43\fdg1, 0\fdg0$). 
The \object{W49} GMC extends over more than 100~pc, assuming a distance of 11.1~kpc \citep{zhg13}. \citet{mad13} derived a total mass 
$M_\mathrm{gas}\sim1.1\times10^6$ M$_\odot$ within a radius 60~pc and $M_\mathrm{gas}\sim2\times 10^5$ $\rm{M_{\odot}}$ within 6~pc from multiscale 
observations of CO. They concluded that the mass reservoir of the molecular cloud is sufficient to form several massive star clusters, or a small system of 
smaller, but still bound, clusters.

The \object{W49} complex consists of two main components; a thermal source (\object{W49A}), which is a star-forming region, and a nonthermal source 
(\objectname[SNR G043.3-00.2]{W49B}), identified as a supernova remnant \citep{mst67}. W49A is one of the most luminous star-forming regions in the Milky Way 
\citep[$L_\mathrm{bol}=10^{7.2}$ L$_\odot$;][]{siv91}, with hundreds of candidate OB stars \citep{dre84,dig90,dep00,alh03}, and consists of the star-forming 
regions W49 north (\object{W49N}/G043.16+0.01), W49 south (\object{W49S}), and W49 southwest (\object{W49SW}). W49N also hosts the most luminous water maser in 
the Galaxy, at a distance of $d=11.4\pm1.2$~kpc \citep{gwn92}. \citet{zhg13} redetermined this distance as $\displaystyle{11.1^{+0.79}_{-0.69}}$~kpc by 
studying the water masers named \objectname[W49 N]{G043.16+0.01} and G048.60+0.02. W49A hosts a ring of a dozen O stars and 25-30 ultracompact \ion{H}{2} (UCHII) 
regions \citep{dre84,wlc87,dig90,dep97}, which may represent a massive star cluster. Its mass has been estimated at $M_\mathrm{cl}$ $\gtrsim$ 4 $\times10^4$ M$_\odot$ \citep{hal05}. 
Recently, \citet{wus14} discovered a very massive O2-3.5 spectral type star in the central cluster of W49A.  

Several authors have tried to explain the nature of the star formation in W49A. \citet{wlc87} postulated a large-scale gravitational collapse toward the 
central ring of hypercompact (HC) \ion{H}{2} regions (Welch ring) in W49N, based on their observations of molecular lines that exhibit a double-peaked profile. 
On the other hand, \citet{ser93} and \citet{bwt96} conclude that this double-peak line profile comes from different clouds and suggest that a cloud$-$cloud collision 
is triggering the massive star formation in W49A. \citet{wil01} found hot cores in the Welch ring, which are probably the precursors of UCHII regions. 
\citet{alh03} identified four massive stellar clusters based on the spatial distributions of the detected sources based on their ($H$-$K_{s}$) colors, and they
hypothesize that the W49 GMC collapsed to form the central massive cluster, and stellar winds and UV radiation triggered the surrounding the cloud to form the 
Welch ring. However, they concluded that there is no evidence of triggering for the other clusters on the south and east parts of the region. In addition, 
\citet{pen10} suggested that the triggering in the region is caused by expanding shells in the center of W49N. Recently, \citet{mad13} studied the mass 
distribution in the whole GMC and concluded that it shows a hierarchical network of filaments at scales from $\sim$ 10 to 100~pc and suggested that the W49A 
starburst is formed from global gravitational contraction. They also concluded the feedback from the central young massive cluster is still not sufficiently 
strong to disrupt the GMC, and there is no evidence for significant disruption from photoionization.

Here we present \textit{Spitzer} IRAC imaging and photometric analysis of the W49 star-forming complex with deep IR data from 1 to 24~$\mu$m to investigate 
the massive YSOs (MYSOs) and embedded clusters forming around them. In Section \ref{sec:obs} we describe the observations, data reduction techniques, and our near- 
and mid-IR source catalog and YSO classification. In Section \ref{sec:analysis} we present the clustering analysis, in Section \ref{sec:sed} we present the 
SED fitting results for massive YSO candidates, in Section \ref{sec:tracers} we present the massive star formation tracers in the region, and in Section \ref{sec:results} we discuss the star formation history in W49 and compare it to other star-forming regions such as G305 and G333. Finally, in Section \ref{sec:sum} we summarize our results and describe our future work.

\begin{figure*}
\centering
\includegraphics[width=15cm]{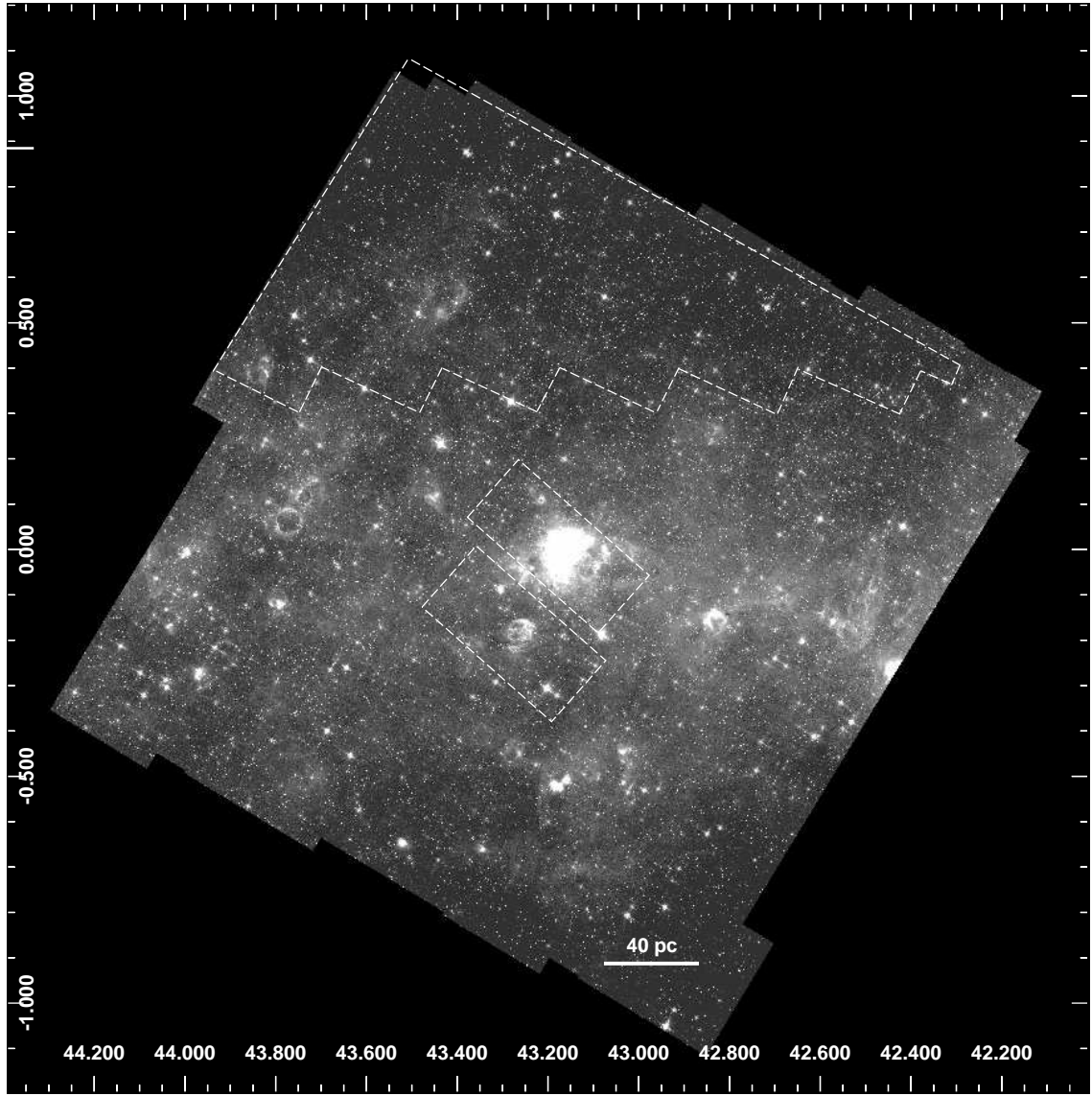}
\caption{5.8~$\mu$m grayscale image of the entire field analyzed in this study. Fields shown by the dashed boxes are where we have the deeper HDR images. The image is centered at \emph{l}, \emph{b} = 43\fdg2, -0\fdg02.}\label{fig:bigmosaic}
\end{figure*}

\section{Observations and Methods} \label{sec:obs}

\subsection{IRAC Imaging} \label{iracimaging}

We assembled the mid-infrared \textit{Spitzer}/IRAC observations of the W49 region obtained with the \textit{Spitzer} IRAC instrument \citep{faz04} at 
$3.6$, $4.5$, $5.8$, and $8.0$~$\mu$m. We list the dates and coordinates of each Astronomical Observation Request (AOR) in Table~\ref{aors}. The data are from 
several projects, including the following project IDs: 631 (PI: G. Fazio), 63 (J. Houck), 187 (GLIMPSE; E. Churchwell), and 80074 (Deep GLIMPSE; B. Whitney).  
The data set contained  a total of 11,592 images with a frame time of 2 s. Also, we used a total of 1088 images acquired in 12 s High Dynamic Range 
(HDR) mode, which performs consecutive individual observations with exposure times of 0.4 and 10.4 s. Before performing the point-source detection and 
photometry, all the IRAC images were processed on an image-by-image basis using the routine imclean,\footnote[6]{See http://irsa.ipac.caltech.edu/data/SPITZER/docs/dataanalysistools/tools/contributed/irac/imclean/} 
which is an IRAF\footnote[7]{IRAF is distributed by the National Optical Astronomical Observatories, operated by the Association of the Universities for Research in Astronomy, Inc., under cooperative agreement with the National Science Foundation.} program for removing the bright source artifacts \citep[``pulldown,'' ``muxbleed,'' and ``banding'';][]{hor04,pip04} from the Basic Calibrated Data (BCD) images.

Automated source detection and aperture photometry were carried out using PhotVis version 1.10 \citep{gut08}. PhotVis utilizes a modified DAOphot \citep{ste87}
source-finding algorithm. Aperture photometry was performed with an aperture of 2.4$\arcsec$ radius and using a background annulus of inner and outer radii 
2.4$\arcsec$, and 7.2$\arcsec$ respectively. Within an area of size $\triangle$\emph{l} $\times$ $\triangle$\emph{b} = 1\fdg68 $\times$ 1\fdg64, centered at 
(\emph{l}, \emph{b}) = (43\fdg2, -0\fdg02), 332,442 sources were detected with IRAC photometry, and among these, 57,254 sources have photometry in all four 
IRAC filters. The photometric catalog is available in the electronic edition of this paper.

\begin{figure*}
\centering
\includegraphics[width=17cm]{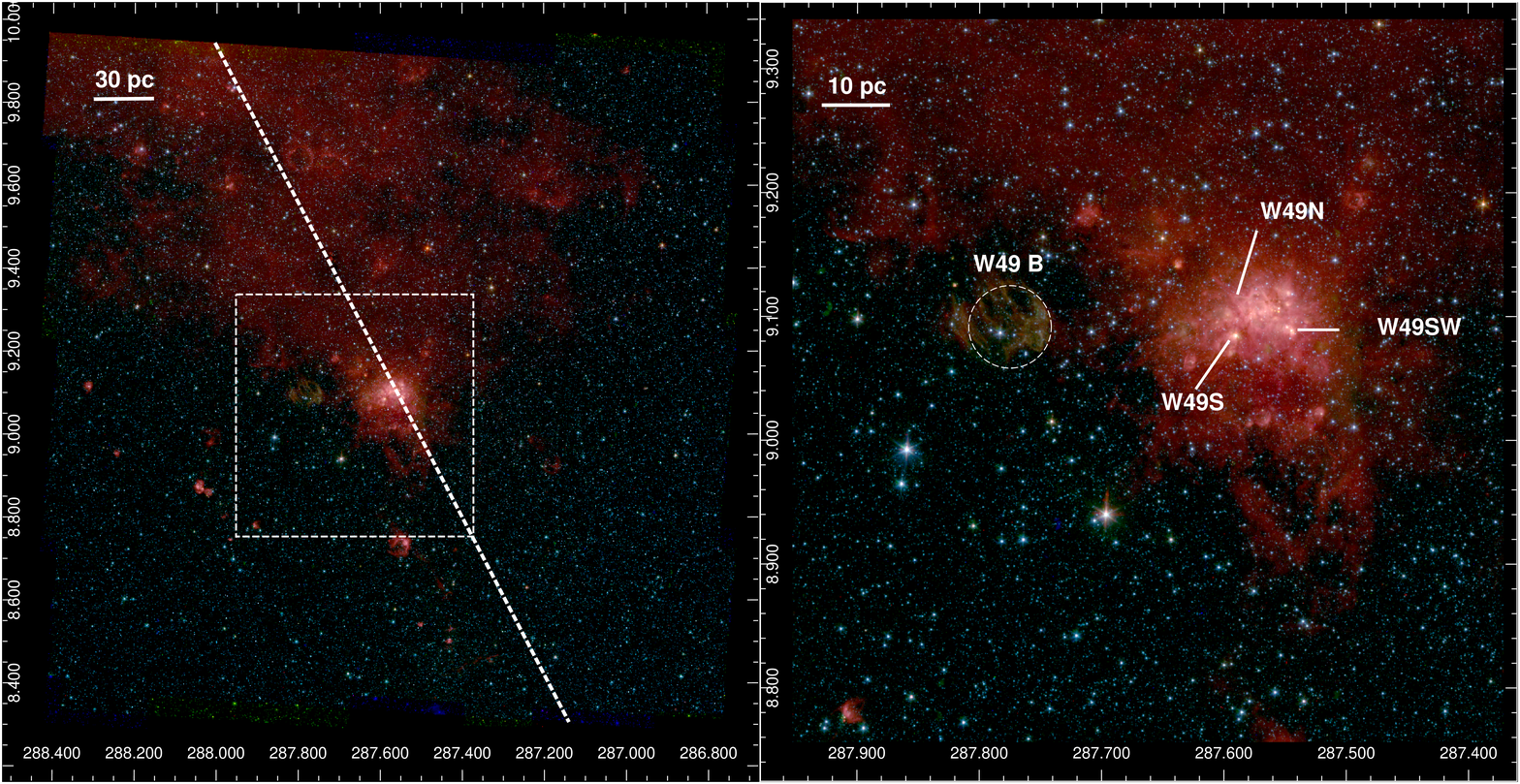}
\caption{The left panel shows a color image of the entire mosaic in the \textit{Spitzer} IRAC bands (blue: 3.6~$\mu$m, green: 4.5~$\mu$m, red: 8.0~$\mu$m).  The dashed line denotes the Galactic plane. The field shown by the dashed gray box is where we performed the clustering analysis, which is shown in detail in the right panel. The size scales in pc assume a distance to W49 of 11.1~kpc. }\label{fig:GMC}
\end{figure*}

The individual BCD images (processed with the \textit{Spitzer} IRAC pipeline version S18.25.0 and S19.1.0) were mosaicked into a larger image using the IRACproc 
package \citep{sch06}. IRACproc is a PDL script based on the Spitzer Science Center's post-BCD processing software MOPEX \citep{mah05}, which has been enhanced 
for better cosmic-ray rejection. The full mosaic corresponds to a region with a size of approximately 360 by 360 parsecs at a distance of 11.1~kpc. 
The $5.8$~$\mu$m grayscale image of the entire mosaic can be seen in Figure~\ref{fig:bigmosaic} and a color image of the GMC with its surroundings can be seen 
in the left panel of Figure~\ref{fig:GMC}. We show the central region W49A in detail and the supernova remnant W49B in the right panel of Figure~\ref{fig:GMC}.

\begin{deluxetable}{lllll}
\tabletypesize{\scriptsize}
\tablecaption{Astronomical Observation Requests \label{aors}}
\tablehead{\colhead{AORKEY} & \colhead{Date} & \colhead{R. A. (J2000)} & \colhead{Decl. (J2000)} & \colhead{IRAC}\\ 
\colhead{ } & \colhead{(UT)} & \colhead{(h m s)} & \colhead{($\degr$ $\arcmin$ $\arcsec$)} & \colhead{Reduction}
\\ \colhead{ } & \colhead{ } & \colhead{} & \colhead{} & \colhead{Pipeline ver.}}
\startdata 
7283968 & 2003 Oct 02 & 19:06:18 & 8:17:36 & S18.25.0\\
4389888 & 2004 Apr 20 & 19:10:14 & 9:06:16 & S18.25.0\\
11963904 & 2004 Oct 09 & 19:11:04 & 9:27:41 & S18.25.0\\
11972096 & 2004 Oct 09 & 19:10:21 & 9:07:30 & S18.25.0\\
11973376 & 2004 Oct 09 & 19:11:47 & 9:47:52 & S18.25.0\\
11966976 & 2004 Oct 09 & 19:09:39 & 8:47:21 & S18.25.0\\
11971072 & 2004 Oct 09 & 19:08:56 & 8:27:04 & S18.25.0\\
11968768 & 2004 Oct 09 & 19:12:30 & 10:08:04 & S18.25.0\\
11972864 & 2004 Oct 09 & 19:13:13 & 10:28:15 & S18.25.0\\
11018240 & 2005 May 06 & 19:11:09 & 9:06:24 & S18.25.0\\
11017984 & 2005 May 06 & 19:11:09 & 9:06:24 & S18.25.0\\
45934848 & 2012 Nov 01 & 19:06:24 & 10:12:04 & S19.1.0\\
45928960 & 2012 Nov 26 & 19:02:31 & 8:20:54 & S19.1.0\\
45949952 & 2012 Nov 28 & 19:03:00 & 8:34:49 & S19.1.0\\
45914880 & 2012 Nov 28 & 19:03:29 & 8:48:43 & S19.1.0\\
45909248 & 2012 Nov 29 & 19:03:58 & 9:02:37 & S19.1.0\\
45940224 & 2012 Nov 30 & 19:05:25 & 9:44:14 & S19.1.0\\
45923328 & 2012 Nov 30 & 19:04:56 & 9:30:22 & S19.1.0\\
45917184 & 2012 Nov 30 & 19:04:27 & 9:16:30 & S19.1.0\\
45945600 & 2012 Nov 30 & 19:09:54 & 10:28:25 & S19.1.0\\
45906176 & 2012 Dec 01 & 19:05:55 & 9:58:09 & S19.1.0\\
\enddata
\end{deluxetable}

\subsection{Completeness Estimate} \label{completenes}

\begin{figure}
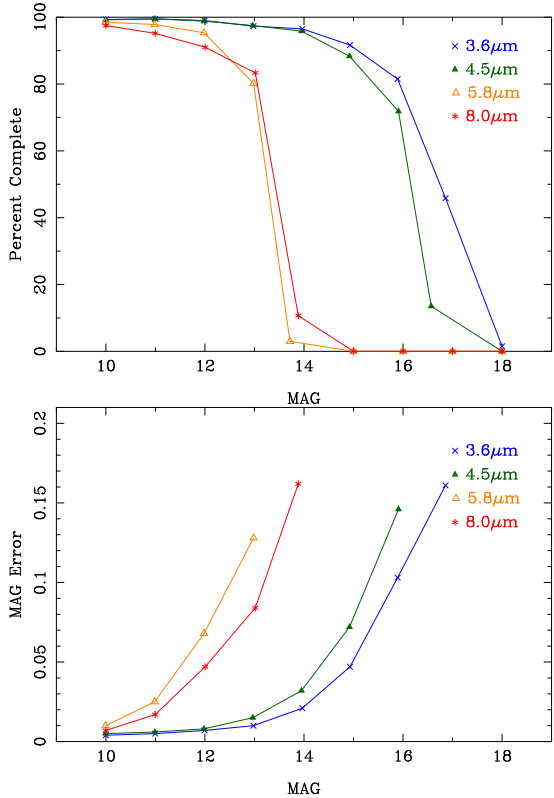

\centering
\includegraphics[angle=270,width=0.4\textwidth]{w49completenessestimate.eps}
\includegraphics[angle=270,width=0.4\textwidth]{w49photometryerrors.eps}
\caption{Top: estimate of the completeness of the IRAC data as a function of magnitude for the survey. Bottom: error estimate as a function of magnitude, derived from the completeness analysis.}\label{fig:completeness}
\end{figure}

\begin{figure}
\centering
\includegraphics[width=8cm]{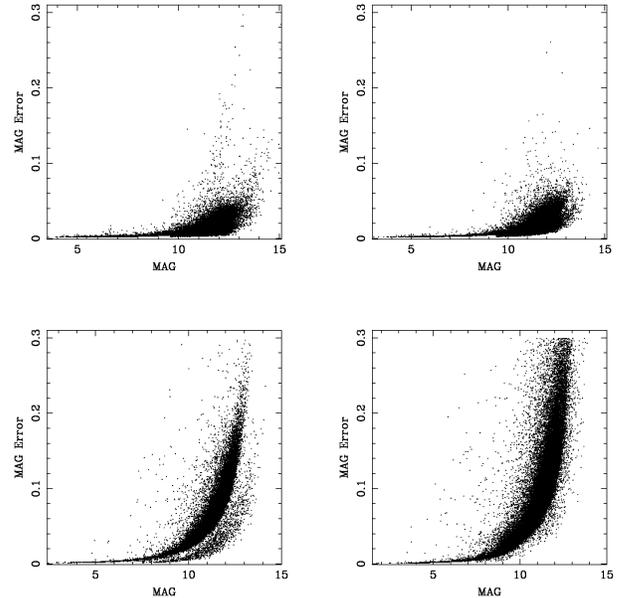}
\caption{Uncertainty in magnitudes versus magnitude in the source catalog for the IRAC 3.6 (top left), 4.5 (top right), 5.8 (bottom left) and 8.0~$\mu$m (bottom right) bands. Sources with errors above 0.3 mag are not plotted.}\label{fig:photerr}
\end{figure}

We estimated the completeness in each IRAC channel by adding artificial sources of various magnitudes to a sample region (centered on 19:10:22, +9:07:35, 1\fdg5x1\fdg5 in size). 
The artificial sources were added to the image using the observed IRAC point-spread function, scaled to the various magnitudes and added in a grid of positions
in the image. We used the same source-finding and photometry routines previously described to extract magnitudes for sources in the field. We then compared 
the result to the input data to determine the completeness percentage and determined the photometric error for the artificial sources. The results of the 
completeness and error estimates are shown in Figure~\ref{fig:completeness}. The $90\%$ completeness magnitudes for IRAC channels 1, 2, 3, and 4 are 15.1, 
14.7, 12.35, and 12.12, respectively. The systematic error of the photometry was determined by the median value of the difference between the calculated 
photometric measurement of the point-spread function (PSF) source and its scaled magnitude and is plotted as a function of magnitude for each of the bands in 
Figure~\ref{fig:completeness}. Figure~\ref{fig:photerr} shows the photometric errors as reported by the PhotVis photometry routine for the catalog sources. 
It can be seen in the 5.8 and 8.0~$\mu$m photometry error plots that the 12 s data provide smaller photometric errors, especially for magnitudes fainter than 12~mag. 

\subsection{Source Catalog} \label{sec:cat}

The source matching for the final catalog was performed in stages. First, all sources detected by \textit{Spitzer} were matched to Two Micron All Sky Survey 
(2MASS) Point Source Catalog \citep{skr06} sources by the PhotVis photometry routine. Since we have catalogs for mosaics based on short (2 s) and long (12 s) 
frames, we ran a band-merging process and took all sources from the short frames brighter than a certain cutoff (magnitude 9.6, 9.4, 7.5, and 7.2 for channels 
1, 2, 3, and 4, respectively), which was chosen to be near but below the saturation limit of the 12 s frames. 

Deep near-IR data are necessary, along with the IRAC photometry, in order to identify YSO candidates from other highly reddened objects and to more accurately 
fit models of their SEDs. Therefore, we used the UKIRT Infrared Deep Sky Survey, DR7PLUS (UKIDSS)\footnote[8]{UKIDSS uses the UKIRT 
Wide Field Camera \citep{cas07} on the United Kingdom Infrared Telescope, and the UKIDSS project is defined by \citet{law07}.} Galactic Plane Survey 
\citep{luc08} data, which are deeper and have better spatial resolution than those of 2MASS. We merged these data into our catalog using the TOPCAT software 
\citep{tay05} pair match method with a maximum 1$\arcsec$ radial tolerance. We adopted UKIDSS data, where they exist, for sources with 2MASS magnitudes fainter
than the UKIDSS saturation limits, and 2MASS data otherwise (the UKIDSS cameras saturate near 12.65, 12.5, and 12 mag in $J$, $H$, and $K_{s}$, respectively). 

UKIDSS data have small uncertainties and small but measurable zero-point photometric offsets from 2MASS. Therefore, we calculated the mean and standard 
deviations of the magnitude residuals between 2MASS and UKIDSS and applied a mean offset to UKIDSS data of 0.02, -0.09, and -0.04 mag in 
$J$, $H$, and $K_{s}$, respectively, to place them on the same system as the 2MASS photometry. Also, we found that the small photometric errors reported in 
the UKIDSS catalog (e.g., $<$0.001 mag for 13 mag sources) are probably unrealistic and have a detrimental effect on the source classification process.
We therefore also adjusted the UKIDSS errors by adding 0.02 mag in quadrature to the values from the UKIDSS catalog before the source classification process. 
This imposes an error floor of 0.02 mag but does not affect the larger errors. We also used the MIPSGAL Archive 24~$\mu$m data \citep{gut15} and added them to 
our catalog by using the TOPCAT software pair match method with a maximum 2$\arcsec$ radial tolerance (since the resolution of the MIPS 24~$\mu$m band is 
6$\arcsec$ FWHM, the position errors can be higher than IRAC). Also, we eliminated the bad matches when they are not identified as point sources in MIPS 
image (see Section \ref{sec:yso}). Our final catalog therefore contains photometry of sources over a wavelength range from 1.2 to 24~$\mu$m. A summary of the 
source catalog and source-matching results can be found in Table~\ref{datasummary} and in Table~\ref{datasourcematch}, respectively. The source catalog itself is presented 
in Table~\ref{sourcetable}.

\begin{deluxetable}{cc}
\tabletypesize{\scriptsize}
\tablecaption{Summary of Source Catalog\label{datasummary}}
\tablewidth{0pt}
\tablehead{\colhead{Band} & \colhead{Number of Sources}}
\startdata
$J$ & 305,584\\ 
$H$ & 321,973\\ 
$K_{s}$ & 322,768\\ 
3.6~$\mu$m & 310,485\\ 
4.5~$\mu$m & 280,892\\ 
5.8~$\mu$m & 93,896\\ 
8.0~$\mu$m & 63,097\\ 
24.0~$\mu$m & 2,268\\          
\enddata
\end{deluxetable}

\begin{deluxetable}{lccc}
\tabletypesize{\scriptsize}
\tablecaption{Source-matching results\label{datasourcematch}}
\tablehead{\colhead{Catalog A} & \colhead{Catalog B} & \colhead{$r_\mathrm{match}$} & \colhead{N. matches}}
\startdata
IRAC & 2MASS & 1.0$\arcsec$ & 221,104 \\
IRAC+2MASS & UKIDSS & 1.0$\arcsec$ & 322,778 \\
IRAC+(2MASS/UKIDSS) & MIPS & 2.0$\arcsec$ & 2268 \\
\enddata
\end{deluxetable}

\subsection{YSO Classification} \label{sec:yso}

In this study we used the selection method based on the color and magnitude criteria defined in \citet{gut09}. This method uses flux ratios or colors to 
identify YSO candidates. This method is an updated version of that introduced by \citet{gut08}. \citet{gut08,gut09} showed that the $K_{s}$-$[3.6]$ vs. $[3.6]-[4.5]$ color$-$color diagram 
is one of the best diagnostics for identifying and classifying YSOs. The $J$ and $H$ bands allow us to further extend the SED and sort out YSO candidates 
from highly reddened main-sequence stars and other background objects. With the near-IR and \textit{Spitzer} bands, we can quickly identify YSO candidates and 
determine their evolutionary status. We classify the YSOs into the categories of Class I sources (protostars with circumstellar disks and infalling envelopes) 
and Class II sources (pre-main-sequence stars with optically thick disks). In addition to these classes, ``deeply embedded sources,'' which are Class I sources 
with bright emission at 24~$\mu$m, and ``transition disks,'' which are Class II sources with significant dust clearing within their disks, can be identified with 
this method.

To identify the YSOs in W49, we first applied the \citet{gut09} criteria to the 332,442 point sources in our catalog. First, we separate out extragalactic 
contaminants such as star-forming galaxies, broad-line active galactic nuclei (AGNs), and polycylic aromatic hydrocarbon (PAH) rich galaxies. Following Gutermuth's criteria, star-forming galaxies are classified from 
very red $5.8$ and $8.0$~$\mu$m colors \citep{ste05} because of their strong PAH feature emission. The $[4.5]-[5.8]$ versus $[5.8]-[8.0]$ and $[3.6]-[5.8]$ 
versus $[4.5]-[8.0]$ color$-$magnitude diagrams and a combination of 8 different conditions based on all four IRAC bands are used to identify the star-forming 
galaxies and PAH galaxies. Broad-line AGNs have mid-IR colors that are largely consistent with YSOs \citep{ste05}. Therefore, the AGNs are classified by using 
the $[4.5]$ versus $[4.5]-[8.0]$ color$-$magnitude diagram and a combination of six different conditions involving these two IRAC bands.

We remove knots of shock emission and sources contaminated by PAH emission using the $[4.5]-[5.8]$ versus $[3.6]-[4.5]$ color$-$color diagram and seven 
different conditions involving these three IRAC bands. Class I YSOs are identified using the same diagram and a combination of two conditions that show their 
discriminant colors. Finally, we extract the Class II YSOs from the remaining objects in the catalog using the $[4.5]-[8.0]$ versus $[3.6]-[5.8]$ color$-$color 
diagram and a combination of four different conditions involving all four IRAC bands.

\begin{figure*}
\centering
\includegraphics[angle=270,width=15cm]{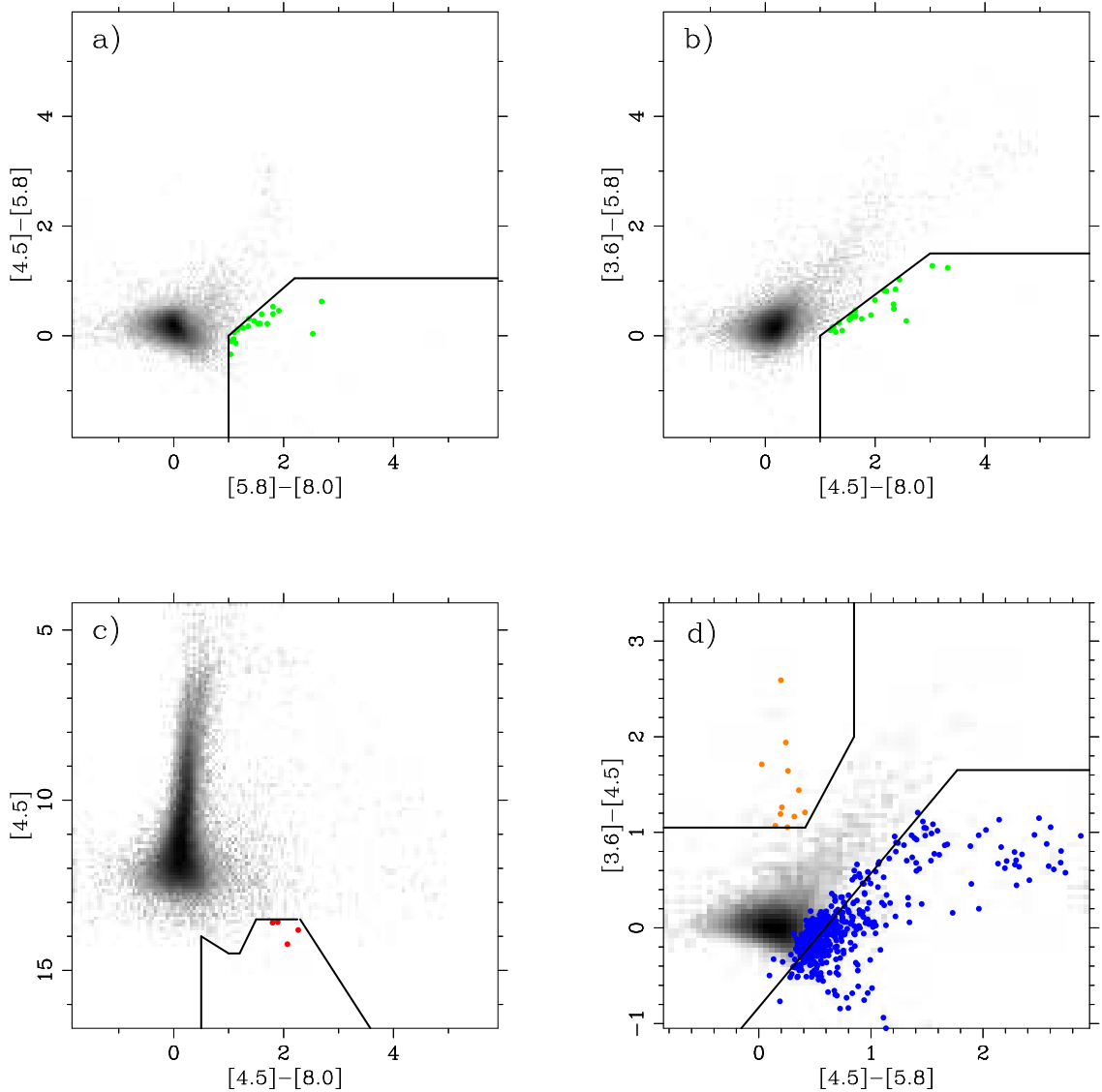}
\caption{Color$-$color and color$-$magnitude diagrams used to identify contaminant objects among the sources with detection at all four IRAC bands following the criteria in \citet{gut09}. 
The background logarithmic gray scale indicates the overall source density in each color$-$color and color$-$magnitude space. In panels (a) and (b), PAH galaxies are marked with green circles. 
In panel (c), candidate AGNs are marked by red circles. Panel (d) shows knots of shocked emission (orange circles) and PAH-contaminated sources (blue circles).}\label{fig:irac4bandconts}
\end{figure*}

With the application of each step of the full Gutermuth contaminant object identification criteria, we selected a total of 4 candidate
broad-line AGNs, 53 PAH-rich galaxies, 11 knots of shocked gas emission, and 639 PAH-contaminated apertures, which are shown in Figure~\ref{fig:irac4bandconts}. 

After removing all these contaminants, the second step in the \citet{gut09} method is to classify the YSOs with near-infrared data, and the third step is to 
reexamine the entire catalog with MIPS 24~$\mu$m photometry. After we applied the full set of criteria, we ended with 73 deeply embedded near-infrared objects, 
271 Class I YSOs, 3021 Class II YSOs, and 231 transitional disk candidates.

In our analysis of the W49 data we had to take into account its greater distance and the possible higher level of contamination from objects such as AGB stars,
background sources, and other extragalactic contaminants that have probably been classified as YSOs. Since AGB stars are bright and, in general, slightly bluer 
than YSOs \citep{rob08,koe14}, we considered the bright YSOs that also follow the following selection criteria as candidate AGB stars:

\begin{displaymath}
3 < [3.6] < 9.5 \text{ and}\ 0.2 < [3.6]-[4.5] < 1.25 
\end{displaymath}
or 
\begin{displaymath}
3.5 < [3.6] < 9.5 \text{ and}\ 0.4 < [3.6]-[8.0] < 2.6
\end{displaymath}

The large cluster of objects at bright magnitudes in $[3.6]-[4.5]$ versus $[3.6]$ and $[3.6]-[8.0]$ versus $[3.6]$ color$-$magnitude diagrams are consistent 
with being AGB stars. With this criterion we classified 212 sources as candidate AGB stars. 

The \citet{gut09} criteria require detections at $5.8$ or $8.0$~$\mu$m to identify extragalactic and background contaminants, and we have only 57,254 sources 
detected in all four IRAC bands. Therefore, for the rest of the sources that have been classified as YSO candidates in Phase II and have only IRAC $3.6$ and 
$4.5$~$\mu$m photometry, we had to think about possible extragalactic contaminants. Thus, we applied a selection cut of [$3.6$] = 13 mag to separate 
YSO candidates from potential background/foreground objects. However, at the distance to W49 that would also remove up to 90$\%$ of the low-mass YSOs that we 
are otherwise sensitive enough to detect. We report these additional YSO candidates separately to denote the lower confidence in their identification but do 
not remove them from our catalog. With this criterion we flagged 64 Class I, 1932 Class II, and 75 transition disk candidates with a ``uc'' mark in our 
catalog, as shown in Table~\ref{sourcetable}, and called them faint YSO candidates. These faint YSO candidate sources will not be used in the analysis described in Section \ref{sec:analysis}.

As a final step, we examined the regions near the YSO candidate objects and eliminated 74 transition disk and 27 embedded source candidates if they did not 
appear as point sources in the MIPS 24~$\mu$m images, seem like an artifact around a very bright source, or are significantly more extended than the FWHM for the 
$24$~$\mu$m PSF.

\begin{figure*}
\centering
\includegraphics[angle=270,width=15cm]{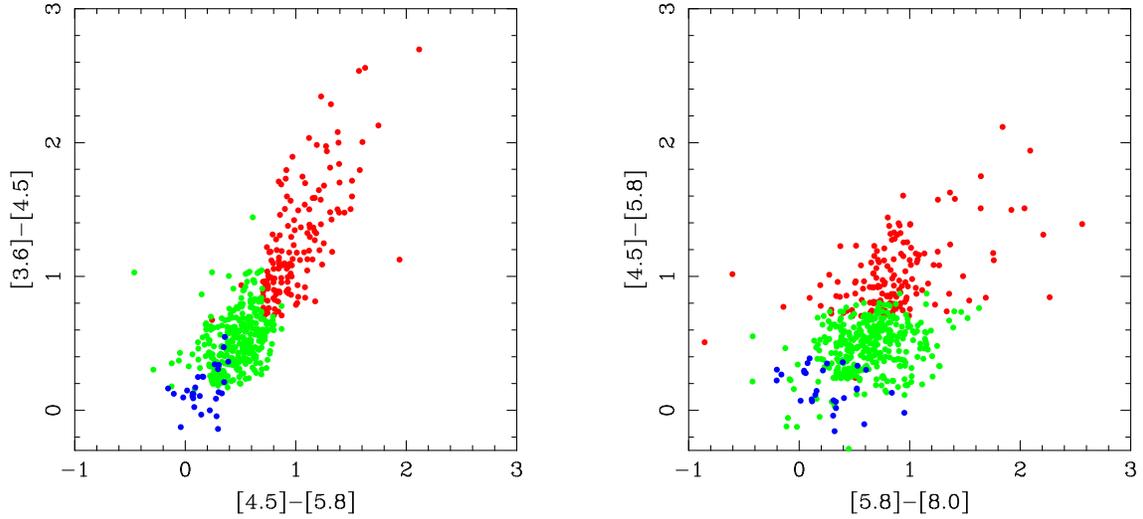}
\caption{IRAC color$-$color diagrams used for identifying YSO candicates in W49. Left: $[3.6]-[4.5]$ vs. $[4.5]-[5.8]$; right: $[4.5]-[5.8]$ vs. $[5.8]-[8.0]$. In both panels, red points are Class I, green points are Class II, blue points are transition disk candidates.}\label{fig:iracysos}
\end{figure*}

\begin{figure*}
\centering
\includegraphics[angle=270,width=15cm]{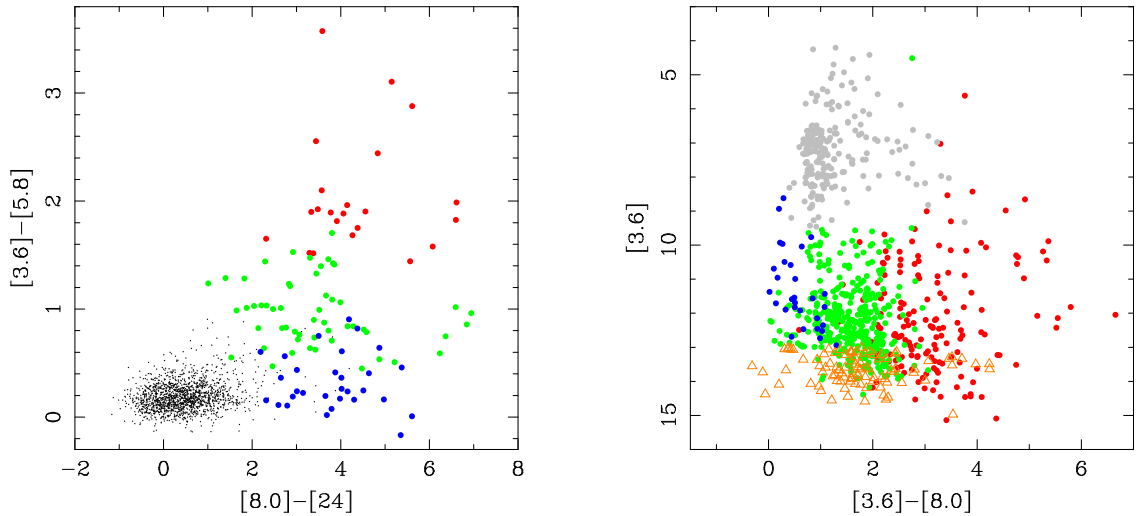}
\caption{IRAC and MIPS color$-$color diagrams used for eliminating the AGB stars and faint objects that could be background/foreground contamination. Left: $[3.6]-[5.8]$ vs. $[8.0]-[24]$; 
right: $[3.6]$ vs. $[3.6]-[8.0]$. The red points are Class I; green points are class II; blue points are transition disk candidates; black points are class III and photospheres; gray points are AGB star 
candidates; and orange points are faint YSO candidates.}\label{fig:IRAC_ClassI_II_III_conts}
\end{figure*}

The color$-$color diagrams of the identified YSOs are shown in Figure~\ref{fig:iracysos} and the color$-$color diagrams combining MIPS and IRAC photometry and 
eliminated AGB star candidates and faint YSOs are shown in Figure~\ref{fig:IRAC_ClassI_II_III_conts}. The points are plotted without dereddening their 
photometry.

We also determined the slope of the SEDs in the mid-IR and examined the resulting class distribution of YSOs. The class identification for the slope of 
log($\lambda$F$_{\lambda}$) vs. log($\lambda$) between $2$ and $\sim20$~$\mu$m from \citet{lad87} is

\begin{equation} \alpha\geq0.3 \textnormal{ Class I}\nonumber\end{equation}
\begin{equation}0.3 >\alpha\geq-0.3 \textnormal{ Flat Spectrum}\nonumber\end{equation}
\begin{equation}-0.3 >\alpha\geq-1.6 \textnormal{ Class II}\nonumber\end{equation}
\begin{equation}-1.6 >\alpha\geq-2.7 \textnormal{ Class III}\nonumber\end{equation}

\begin{figure}
\centering
\includegraphics[width=8cm]{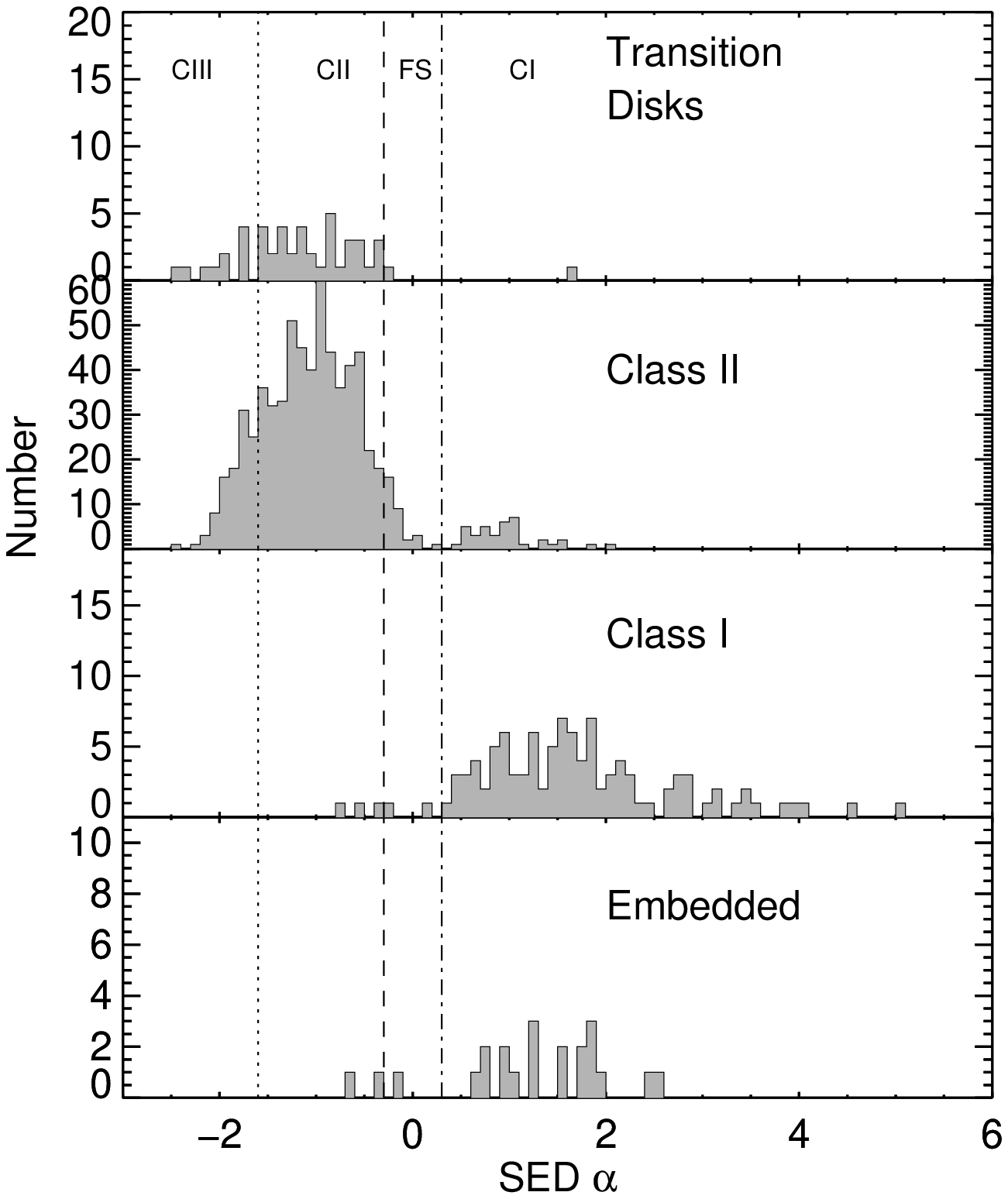}
\caption{Distribution of the SED spectral index $\alpha$ for sources identified as YSO candidates. The three vertical lines mark the division between the YSO regions based on their SED slopes. From left to right: Class II, flat spectrum, and Class I.}\label{fig:sedslopes}
\end{figure}

In Figure~\ref{fig:sedslopes}, we show the identified YSO candidates with the distribution of their SED spectral index. The embedded sources have mainly positive spectral slopes, similar to Class I YSOs. Most of the IRAC-identified
Class I YSO candidates are found to have positive spectral index, indicating substantial infrared excess and confirming their identification as early stage 
YSOs. The IRAC-identified Class II YSOs are mainly grouped in the CII region as shown in Figure~\ref{fig:sedslopes}, consistent with the classification according to the \citet{gut09} method.  The slope distribution for transition disk candidates agrees with these sources being evolved YSOs. 

The full list of objects in Table~\ref{sourcetable} includes their classification (if available) and SED spectral index. In Table~\ref{sourcesum} we summarize 
the final source classification results.

\begin{deluxetable}{lr}
\tabletypesize{\scriptsize}
\tablecaption{Source Classification Summary\label{sourcesum}}
\tablewidth{0pt}
\tablehead{\colhead{Class} & \colhead{Number of Objects}} 
\startdata
Class I & 186\\
Class II & 907\\
Transition disks & 74\\
Embedded protostars & 46\\
Faint Class I\tablenotemark{*}& 64\\
Faint Class II\tablenotemark{*}& 1932\\
Faint transition disk\tablenotemark{*}& 75\\
Photospheres & 247,142\\
AGB candidates & 212\\
Other \tablenotemark{**} & 705\\
Unclassified & 81,097\\
Total & 332,442\\
\enddata
\tablenotetext{*}{These are the YSO candidates with lower confidence.}
\tablenotetext{**}{Includes PAH-emission dominated sources,
  shocked-gas-emission-dominated sources, broad-line AGN
  candidates, and PAH galaxy candidates. Unclassified sources lack detection in four
  bands (either $HK_S$, IRAC 1 and 2, or IRAC 1, 2, 3 and 4) or a
  bright MIPS 24~$\micron$ detection.}
\end{deluxetable} 

\subsection{Comparison with WISE}
We assembled the \textit{WISE} data for the W49 region from the AllWISE Data Release \citep{cut13} and used the YSO selection method that is based on color and 
magnitude criteria defined in \citet{koe14}. We compared these data with the \textit{Spitzer} catalog using the source positions, considering objects a match if the 
distance between the \textit{WISE} and \textit{Spitzer} sources was less than 6$\arcsec$. Nearly half of the IRAC YSOs did not have any match in the \textit{WISE} point-source catalog. 
Most of the IRAC YSOs that did not have corresponding entries in the \textit{WISE} catalog are faint or unresolved in \textit{WISE}. This includes the central portion of W49, 
where the source density is too high given the resolution of \textit{WISE} to allow point-source detection among the bright nebular backgrounds. In the small number of 
overlapping sources, there are no obvious trends that explain all of the differences in classification. Overall, \textit{WISE} channel 1 measurements are a bit dimmer 
for all the YSOs than the IRAC channel 1 measurements. The IRAC channel 2 vs. \textit{WISE} channel 2 has a lot of scatter (differences up to about 0.5 mag), 
but they look more centered around 0. The correspondence between the longer \textit{WISE} and \textit{Spitzer} wavelengths is not as close, but the YSOs appear much brighter 
at long wavelengths according to \textit{WISE} than the \textit{Spitzer} values. This effect could be due to the larger \textit{WISE} beam and other sources or extended emission being 
included in the \textit{WISE} photometry of the YSOs.

\begin{figure}
\centering
\includegraphics[width=8cm]{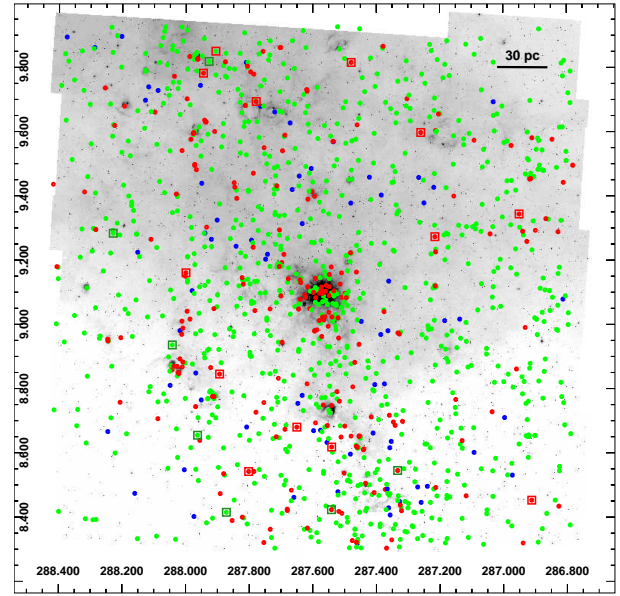}
\caption{IRAC and \textit{WISE} YSO candidate distribution overlaid on the IRAC 8~$\mu$m image. The IRAC YSOs are shown with the filled circles, with red for Class 0/I, green for Class II, blue for transition disks. The \textit{WISE} YSO candidates are plotted as boxes, with red for \textit{WISE} Class I and green for \textit{WISE} Class II.}\label{fig:sdist}
\end{figure}

We found that a total of 47 \textit{WISE} YSOs (25 Class I and 22 Class II) match to IRAC point sources. However, only 20 of them match to the IRAC YSOs within 
50$\arcsec$ of the center of W49, shown in Figure~\ref{fig:sdist}. Within this group of 20 \textit{WISE} YSOs, 12 Class I candidates match IRAC Class I candidates, 
1 \textit{WISE} Class I candidate matches an IRAC Class II candidate, 5 \textit{WISE} Class II candidates match IRAC Class II candidates, and 2 \textit{WISE} Class II candidates match 
IRAC Class I candidates. The rest of the \textit{WISE} YSO candidates (27 sources) match to IRAC AGB stars, photospheres, or PAH-contaminated source candidates. 
The list of \textit{WISE} YSOs that match with IRAC YSOs is given in Table~\ref{WISEYSOs} and the sources are shown on an $8.0$~$\mu$m grayscale image in 
Figure~\ref{fig:sdist}.

The source classification with \textit{WISE} data indicates that the high density of sources in the central region, saturation, and source confusion prevent us from identifying
YSO candidates. Differences in classification may also originate from the inaccurate photometry and lead us to identify some contaminant sources as YSO candidates with the \textit{WISE}
data. Because of these effects, we conclude that the \textit{WISE} data are not very useful at identifying YSO candidates in the central part of W49, and therefore we did not include \textit{WISE} YSO candidates in our further analysis. This conclusion may apply to other regions that are similarly distant or have a high density of sources.  

\clearpage
\begin{turnpage}
\begin{deluxetable}{lcccccccccccccc}
\tabletypesize{\scriptsize}
\tablecaption{Source List\label{sourcetable}}
\tablewidth{0pt}
\tablehead{\colhead{Source Name\tablenotemark{a}} & \colhead{R.A.} & \colhead{Decl.} & \colhead{$J$} & \colhead{$H$} & \colhead{$K_{s}$} & \colhead{$[3.6]$} & \colhead{$[4.5]$} & \colhead{$[5.8]$} & \colhead{$[8.0]$} & \colhead{$[24]$} & \colhead{} & \colhead{} & \colhead{} & \colhead{SED}\\
\colhead{} & \colhead{J2000.0} & \colhead{J2000.0} & \colhead{(mag)} & \colhead{(mag)} & \colhead{(mag)} & \colhead{(mag)} & \colhead{(mag)} & \colhead{(mag)} & \colhead{(mag)} & \colhead{(mag)} & \colhead{Type\tablenotemark{b}} & \colhead{Phase\tablenotemark{c}} & \colhead{Flag\tablenotemark{d}} & \colhead{slope $\alpha$}\\
\colhead{} & \colhead{($\degr$)} & \colhead{($\degr$)} & \colhead{} & \colhead{} & \colhead{} & \colhead{} & \colhead{} & \colhead{} & \colhead{} & \colhead{} & \colhead{} & \colhead{} & \colhead{} & \colhead{}}
\startdata
SSTOERC$\_$G043.0997-00.0296 & 287.56134 & 9.02556 & 19.74(15) & 17.04(03) & 15.54(03) & \nodata & 12.27(08) & \nodata & \nodata & 4.04(10) & 0 & 0 & \nodata & 1.51 \\
SSTOERC$\_$G042.6828+00.2115 & 287.15012 & 8.76694 & 17.13(02) & 16.03(02) & 14.90(02) & 11.56(02) & 10.37(01) & 9.04(02) & 8.18(02) & \nodata & 1 & 1 & \nodata & 1.88 \\
SSTOERC$\_$G042.2532+00.1846 & 286.97418 & 8.37315 & 16.44(02) & 14.80(02) & 14.08(02) & 13.03(05) & 12.44(05) & 11.70(09) & 10.95(11) & 7.5(09) & 2 & 1 & \nodata & -0.47 \\
SSTOERC$\_$G042.8060+00.6517 & 286.81198 & 9.07888 & 13.95(02) & 13.08(02) & 12.68(02) & 12.35(03) & 12.21(02) & 11.91(10) & 11.31(11) & 8.30(09) & 3 & 1 & \nodata & -1.72 \\
SSTOERC$\_$G043.2857$-$00.5262 & 288.09421 & 8.96086 & 18.83(06) & 17.12(04) & 16.10(03) & 14.62(14) & 13.71(09) & \nodata & \nodata & \nodata & 1 & 3 & uc & 0.20 \\
SSTOERC$\_$G043.1518$-$00.9278 & 288.39154 & 8.65611 & 15.53(02) & 14.94(02) & 14.31(02) & 13.93(06) & 13.20(18) & \nodata & \nodata & \nodata & 2 & 3 & uc & -1.93 \\
SSTOERC$\_$G042.4240+00.1903 & 287.04852 & 8.52740 & 15.81(02) & 14.59(02) & 13.96(02) & 13.35(07) & 13.36(09) & \nodata & \nodata & 8.27(10) & 3 & 3 & uc & -0.87 \\  
SSTOERC$\_$G042.5060+00.5319 & 286.78003 & 8.75746 & 17.55(01) & 15.56(02) & 14.67(02) & 13.84(03) & 13.91(03) & 13.44(11) & \nodata & \nodata & 99 & 1 & \nodata & \nodata \\
SSTOERC$\_$G042.7282+00.4616 & 286.94656 & 8.92236 & \nodata & 18.70(14) & 17.30(09) & 14.55(13) & 13.64(08) & 12.40(13) & 11.55(12) & \nodata & 29 & 0 & \nodata & \nodata \\
SSTOERC$\_$G042.9654+00.6453 & 286.89200 & 9.21747 & 15.85(02) & 13.88(02) & 13.02(02) & 12.33(01) & 12.43(02) & 10.51(05) & 12.18(13) & \nodata & 19 & 0 & \nodata & \nodata \\
SSTOERC$\_$G042.8351+00.3941 & 287.05713 & 8.98614 & 5.62(02) & 5.09(05) & 4.88(03) & 6.64(01) & 4.93(00) & 4.90(00) & 4.85(00) & \nodata & 9 & 0 & \nodata  & \nodata \\
SSTOERC$\_$G042.8251+00.6911 & 286.78540 & 9.11397 & 14.91(02) & 12.76(02) & 11.83(02) & 11.24(01) & 11.43(01) & 11.03(05) & 11.04(06) & \nodata & 20 & 0 & \nodata  &\nodata\\
SSTOERC$\_$G043.1583+00.8687 & 286.78085 & 9.49150 & 17.29(03) & 13.29(02) & 10.26(02) & 6.48(00) & 5.75(01) & 3.87(00) & \nodata & \nodata & 12 & 3 & \nodata &\nodata \\
SSTOERC$\_$G042.8033+00.6849 & 286.78085 & 9.09181 & 17.43(03) & 16.15(03) & 15.60(03) & 14.79(12) & 14.28(12) & \nodata & \nodata & \nodata & 4 & 3 & \nodata & \nodata\\
\enddata
\tablecomments{Table 5 is published in the electronic edition of the {\it Astrophysical Journal}. A portion is shown here for guidance regarding its form and content. Values in parentheses signify error in last 2 digits of magnitude value. Right ascension and Declination coordinates are J2000.0.} 
\tablenotetext{a}{Sources are named according to their Galactic longitude and latitude with the prefix SSTOERC, referring to Spitzer Space Telescope, Origin and Evolution of Rich Clusters project.}
\tablenotetext{b}{0; deeply embedded, 1; Class I, 2; Class II, 3; transition disks, 99; Class III and photospheres, 9; shocked gas emission, 29; AGNs, 19; PAH galaxies, 20; PAH dominated sources, 12; AGBs, -100; unclassified.}
\tablenotetext{c}{Sources are flagged according to the phase they were classified.1; \citet{gut09} classification Phase I, 2; Phase II, 0; Phase III, 3; our AGBs, faint YSO/background contamination phase (see Section 2.4).}
\tablenotetext{d}{YSO candidates near a bright IRAC source within 35$\arcsec$ are flagged with a question mark. Faint YSOs which can be background/foreground contamination are flagged with the letter ``uc''.}
\end{deluxetable}
\end{turnpage}
\clearpage

\begin{deluxetable*}{llrrrrrlll}
\tabletypesize{\scriptsize}
\tablecaption{\textit{WISE} YSO Source List\label{WISEYSOs}}
\tablewidth{0pt}
\tablehead{\colhead{\textit{WISE} R.A.} & \colhead{\textit{WISE} Decl.} & \colhead{$W1$} & \colhead{$W2$} & \colhead{$W3$} & \colhead{$W4$} & \colhead{\textit{WISE}} & \colhead{Matched IRAC} & \colhead{IRAC}\\
\colhead{($\degr$)} & \colhead{($\degr$)} & \colhead{(mag)} & \colhead{(mag)} & \colhead{(mag)} & \colhead{(mag)} & \colhead{Class\tablenotemark{a}} & \colhead{Source Name} & \colhead{Class\tablenotemark{a}} }
\startdata
287.2152079	&	9.2737751	&	14.05(21)	&	12.36(07)	&	8.90(04)	&	5.99(06)	&	1	&	SSTOERC G043.1622+00.3880	&	1	\\
287.4789811	&	9.8163322	&	12.33(06)	&	9.93(03)	&	5.45(02)	&	2.05(02)	&	1	&	SSTOERC G043.7637+00.4076	&	1	\\
287.2595223	&	9.5982765	&	13.05(07)	&	11.38(04)	&	8.83(07)	&	5.70(05)	&	1	&	SSTOERC G043.4703+00.4988	&	1	\\
286.9496205	&	9.3442128	&	11.55(09)	&	10.27(02)	&	6.91(02)	&	3.95(02)	&	1	&	SSTOERC G043.1041+00.6530	&	1	\\
287.8935132	&	8.8461252	&	10.48(03)	&	9.07(02)	&	6.72(04)	&	4.48(05)	&	1	&	SSTOERC G043.0923\textminus00.4037	&	1	\\
287.9997854	&	9.1612432	&	11.48(03)	&	9.66(02)	&	7.24(04)	&	4.84(03)	&	1	&	SSTOERC G043.4202\textminus00.3508	&	1	\\
287.7786347	&	9.6947734	&	11.65(05)	&	9.5(03)	        &	5.90(02)	&	2.43(02)	&	1	&	SSTOERC G043.7923+00.0895	&	1	\\
287.9451674	&	9.7828376	&	12.20(05)	&	10.68(03)	&	7.48(05)	&	3.98(05)	&	1	&	SSTOERC G043.9462\textminus00.0153	&	1	\\
287.5414301	&	8.4239783	&	11.63(03)	&	10.60(03)	&	8.12(05)	&	5.34(04)	&	2	&	SSTOERC G042.5570\textminus00.2899	&	1	\\
287.5407128	&	8.6191474	&	11.04(03)	&	9.32(02)	&	6.11(02)	&	3.85(03)	&	1	&	SSTOERC G042.7298\textminus00.1991	&	1	\\
287.3331588	&	8.5458962	&	10.43(02)	&	9.22(02)	&	7.06(03)	&	5.73(08)	&	2	&	SSTOERC G042.5701\textminus00.0509	&	1	\\
286.9115172	&	8.4531359	&	11.19(04)	&	9.62(02)	&	7.06(02)	&	4.80(05)	&	1	&	SSTOERC G042.2956+00.2765	&	1	\\
287.6502571	&	8.6810581	&	9.53(02)	&	6.61(02)	&	4.13(01)	&	1.76(02)	&	1	&	SSTOERC G042.8347\textminus00.2668	&	1	\\
287.8018156	&	8.5427571	&	9.17(02)	&	7.29(02)	&	5.20(01)	&	3.90(02)	&	1	&	SSTOERC G042.7814\textminus00.4635	&	1	\\
288.0424475	&	8.9366108	&	9.34(02)	&	8.89(02)	&	7.68(03)	&	5.34(06)	&	2	&	SSTOERC G043.2406\textminus00.4921	&	2	\\
287.9274058	&	9.8193046	&	9.70(02)	&	8.85(02)	&	6.37(02)	&	4.02(03)	&	2	&	SSTOERC G043.9705+00.0170	&	2	\\
287.9060792	&	9.8514046	&	11.08(03)	&	9.68(02)	&	7.47(04)	&	4.74(06)	&	1	&	SSTOERC G043.9892+00.0504	&	2	\\
288.2291642	&	9.2840378	&	11.59(04)	&	10.46(04)	&	8.25(05)	&	5.49(05)	&	2	&	SSTOERC G043.6338\textminus00.4946	&	2	\\
287.8718301	&	8.4157962	&	11.98(03)	&	11.13(03)	&	8.81(06)	&	6.18(06)	&	2	&	SSTOERC G042.7008\textminus00.5836	&	2	\\
287.9626502	&	8.6556335	&	12.02(04)	&	11.61(04)	&	9.01(09)	&	5.90(07)	&	2	&	SSTOERC G042.9550\textminus00.5526	&	2	\\
\enddata
\tablenotetext{a}{0: deeply embedded; 1: Class I; 2: Class II candidate sources.}
\end{deluxetable*}

\section{Clustering Analysis} \label{sec:analysis}

\subsection{Spatial Distribution of YSOs} \label{sec:spt}

Figure~\ref{fig:sdist} shows the spatial distribution of the YSOs overlaid on the $8.0$~$\mu$m grayscale image of the region. The sources are colored 
according to their evolutionary class: blue for transition disks, green for Class II, and red for Class 0/I candidates. In the central W49A region, it is clear 
that Class 0/I sources and Class II sources are clustered together. In the whole region, however, the Class 0/I sources are clustered, or in filamentary form, 
while the Class II sources are more evenly distributed.

\subsection{Minimal Spanning Tree (MST)} \label{sec:mstmethod}

To understand the star formation history in W49 and compare W49 to other star-forming regions, we want to identify individual groups or clusters of YSOs and 
compare their properties such as sizes and ages. In order to find the clusters, we used a method called MST \citep{caw04}, which uses 
the spatial distribution of sources to determine a cluster membership without any kinematic information.

The MST method defines clusters as a collection of stars that are connected to each other by branches smaller than the cutoff distance ($d_{c}$) or branch 
length and with a minimum number of stars ($N$) in a group. It has been a popular tool in recent years and has been used for many nearby star-forming regions 
\citep{koe08,gut09,bee10,bil11,cha14}. However, there is no unique way to determine a cutoff distance for a cluster determination. 

The most common way used for nearby regions is to plot the distribution of cutoff distances, fit straight lines through the long and short cutoff distance 
domains, and choose the point of intersection \citep{gut08}. Choosing a cutoff distance that falls between these two domains is also a good way to separate 
clusters from distributed sources \citep{gut08,gut09}. With this method, \citet{bee10} determined the cutoff distance as 61$\arcsec$ (0.43~pc) for the 
Diamond Ring region in the Cygnus-X star-forming complex, which is at a distance of 1450~pc. On the other hand, \citet{gua13} derived the cutoff distance as 
72$\arcsec$ (0.51~pc) for the Cygnus-OB2 region in the Cygnus-X complex. All these values are smaller than the 88$\arcsec$ (0.86~pc) derived by \citet{koe08} 
for the W5 star-forming region, which lies at a distance of 2~kpc, which may indicate that Cygnus-X is more densely populated with YSOs than W5. 

\citet{cha14} used the MST method for nearby embedded star-forming regions by dividing each region according to their YSO concentrations. For W5-east they 
found an average cutoff distance of 33$\arcsec$ (0.32~pc), which is different than the 88$\arcsec$ (0.86~pc) that was found by \citet{koe08} for the whole of W5. 
\citet{bil11} used the MST method according to the definition given in \citet{gut09} for Galactic star-forming regions like W43 and VulOB1 to derive their 
clustering properties and found cutoff distances of $\sim$80$\arcsec$ (corresponding to 4.3~pc at the distance of W43) and $\sim$100$\arcsec$ (5.4~pc), 
respectively. These cutoff distance differences may be partly a resolution effect, since at greater distances multiple stars may blend into single objects, 
and fewer stars are detected owing to corresponding lower sensitivity in the more distant regions. Low-resolution observations of distant star-forming regions 
probe only larger-scale structures, and it is hard to see the hierarchical substructure in these regions. In addition to Galactic regions, the MST method has 
been used by \citet{bas07} for the star-forming regions in the M33 galaxy. In order to prevent an artificial length scale according to the resolution limit and
seeing a large number of stars as blended or as single sources, they used cutoff distances from 5$\arcsec$ to 65$\arcsec$ (corresponding to 19 to 252~pc assuming 
the distance of M33 as 800~kpc) in 13 equal steps. Examining the clustering at the different values allowed them to see all levels of hierarchy in M33.

\subsubsection{W49 Clustering Analysis with MST Method} \label{sec:mstw49}

To compare with previous clustering analysis in nearby star-forming regions, we first use the straight-line fit method to determine the branch length cutoff for W49. 
We plot the number of clustered objects versus branch length. Branch lengths between 82$\arcsec$ and 130$\arcsec$ deviated significantly from linearity, 
so we masked these areas and found the intersection to the best fits to the long and short branch length regimes. We found a cutoff distance of 96$\arcsec$,
which corresponds to 5.2~pc at 11.1~kpc, similar to the cutoff distance found by \citet{bil11} for W43 and VulOB1. However, this cutoff distance shows that the straight-line fit
method does not find the isolated subclusters, instead picking up almost the whole star-forming region with a diameter of 27~pc associated with the overdensity of YSOs.

To determine the statistical significance of the W49 clusters identified with this cutoff distance, we performed simulations to examine the clustering properties of randomly 
distributed objects. We created 1000 distributions of 231 objects spread randomly in a 0.5 x 0.5 degree field to match the number of YSOs
and the size of the central region of W49. We examined the number of groups that would occur in a random distribution as a function of the minimum group size. We chose the 
minimum group size that gave the fewest random groups for a given branch length.

We found that for a cutoff distance of 96$\arcsec$ and a minimum group size $N$~$=$~7, our random distribution of YSOs will yield on average nine groups or subclusters. 
We looked at the group with the highest number of members in each simulation and found that these largest groups had an average of 22$\pm$6 members, with a maximum of 78.
For our observed distribution we see a much smaller number of clusters, finding three concentrated groups of YSOs containing 52\% of the observed sources using a 96 $\arcsec$ 
cutoff distance. The population of the largest observed group (Cluster 1, G43.15$-$0.01) is 97, significantly higher than the maximum seen in simulated random distributions. 
Cluster 2 (G43.33$-$0.08) has 15 members, similar to the 16$\pm$2.9 average size of observed for random distributions of objects. Cluster 3 (G43.31$-$0.20) has nine
members, slightly smaller than the 13$\pm$2.0 found for the third-largest random subcluster. The remaining 110 YSOs (48\%) did not belong to any of these clusters and are considered to be part of the distributed population.

\begin{figure*}
\includegraphics[width=8cm]{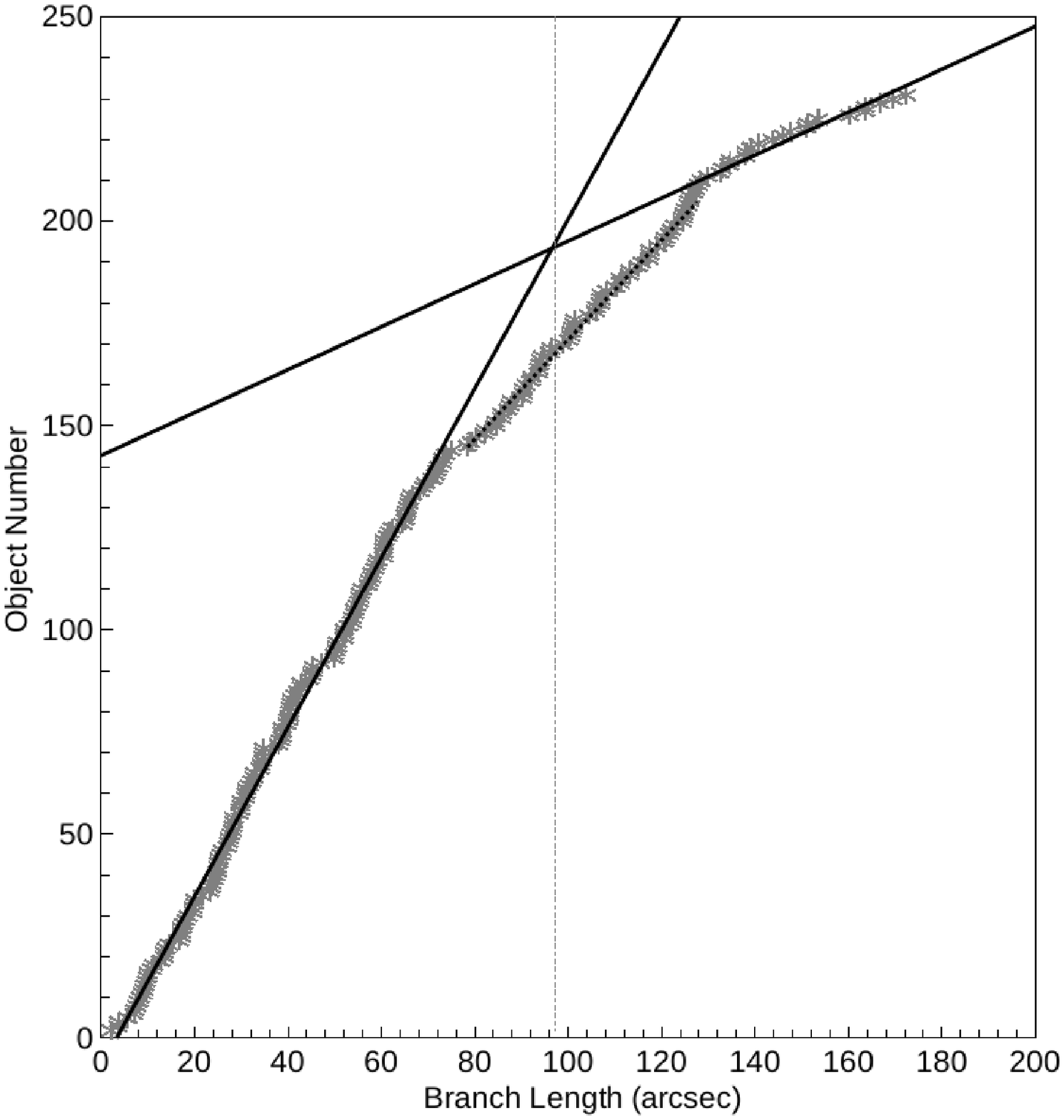}
\includegraphics[width=8cm]{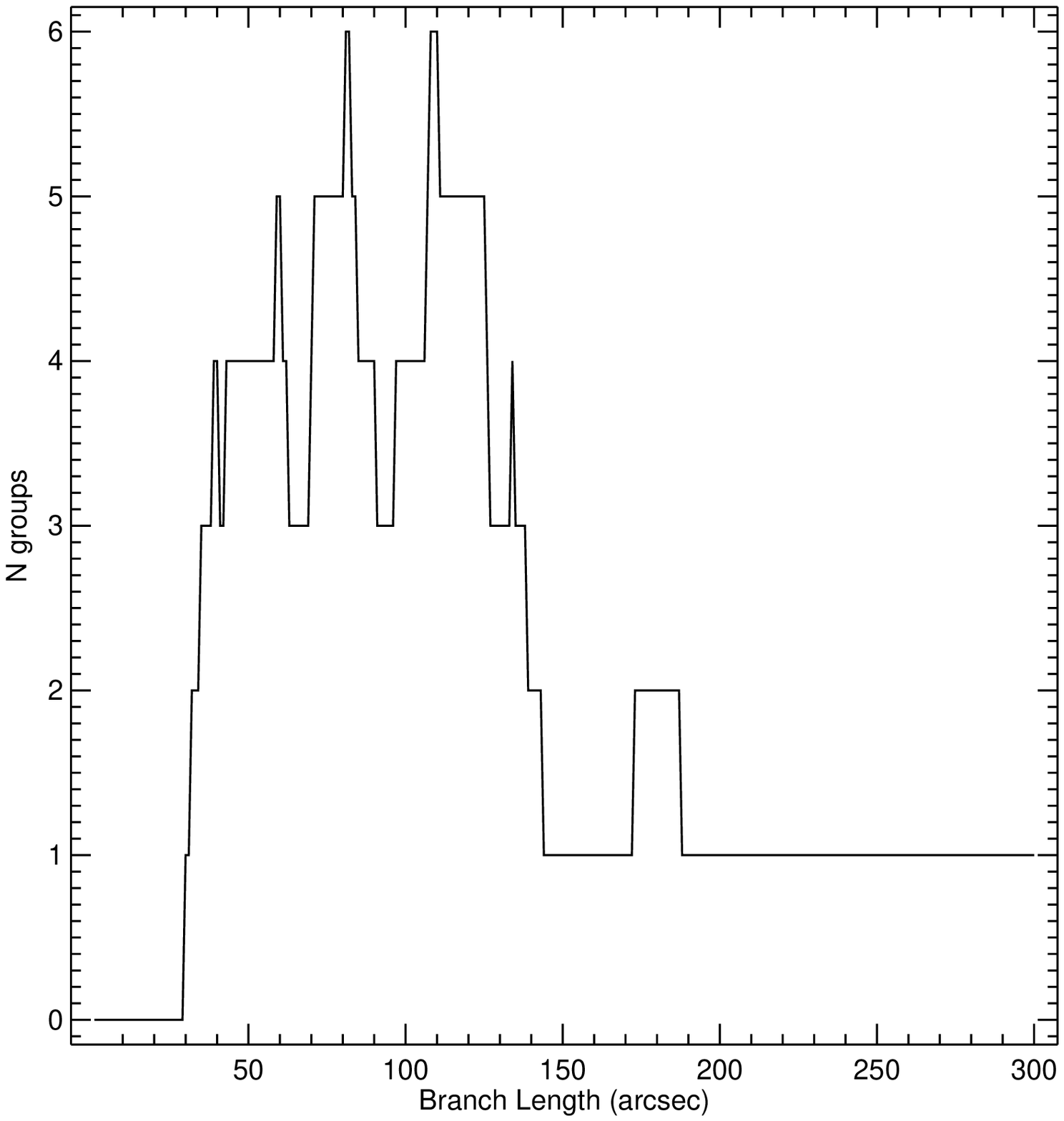}
\caption{Left: The MST branch length distribution. The straight lines represent the linear fit to the points smaller and larger than the chosen branch length. The point of intersection \emph{d$_c$} = 96$\arcsec$, and the minimum number of stars $N$ = 7. Right: Number of groups containing seven or more stars identified by the MST algorithm.\label{fig:mst}}
\end{figure*}

\begin{figure*}
\includegraphics[width=8.2cm]{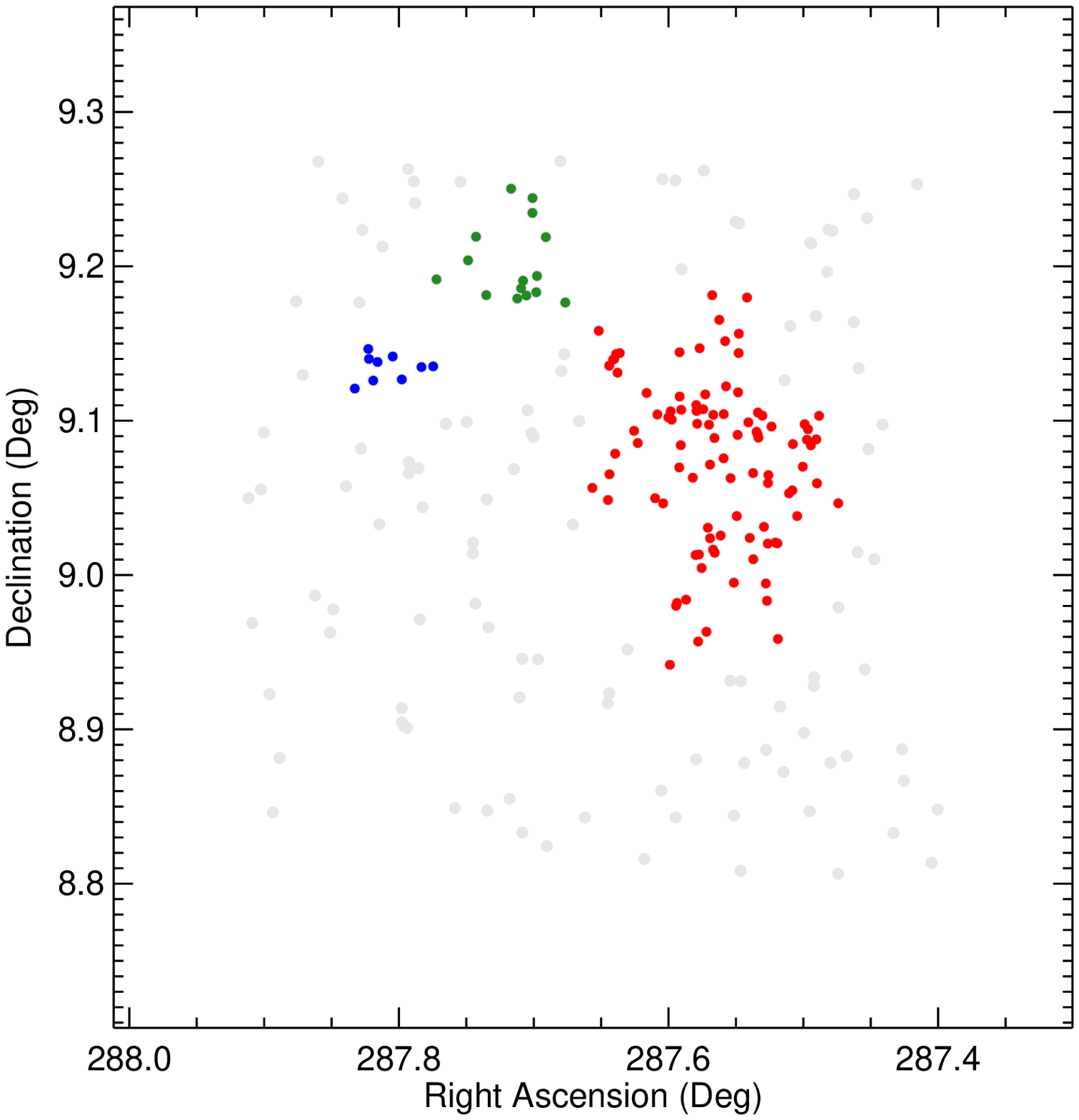}
\includegraphics[width=8.2cm]{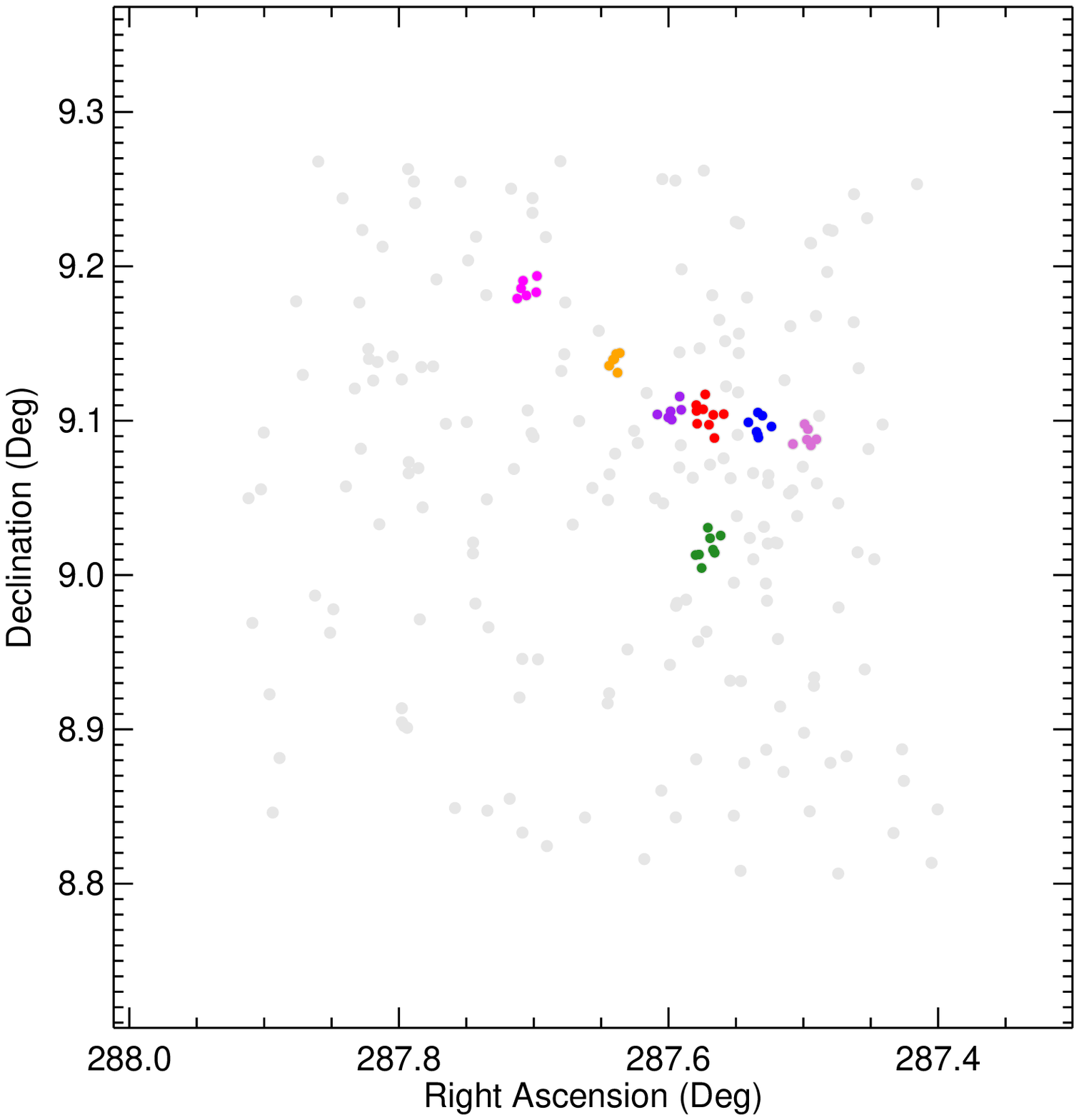}
\caption{Left: three MST clusters in the W49 determined using a cutoff distance of  96$\arcsec$ (5.2~pc for objects at the distance of W49). Right: seven MST subclusters identified with $N$ $\geq$ 6 and $d_c$ = 40$\arcsec$ (2.2~pc). Objects in clusters are colored according to their group, while sources not associated with any clusters are plotted in gray.}\label{fig:mstresults1}
\end{figure*}

The identified clusters in W49 are shown in the left panel of Figure~\ref{fig:mstresults1} and their properties are summarized in Table~\ref{GMCclusters1}. 
The distribution of cutoff distances versus the number of sources is plotted in the left panel of Figure~\ref{fig:mst}. To see how the numbers of groups 
change according to the cutoff distance, we plotted the number of groups containing seven or more stars versus cutoff distance from 1$\arcsec$ to 300$\arcsec$,
with steps of 1$\arcsec$, in the right panel of Figure~\ref{fig:mst}.

This comparison indicates that the largest group observed, cluster 1, is likely formed from the extended YSO population of W49, representing 
a large-scale structure extending 26.86~pc in diameter. The lower-density clusters 2 and 3 observed with the straight-line fit cutoff distance of 96 $\arcsec$ 
have a higher probability of occurring randomly as a result of chance alignment of distributed YSOs.

To investigate the substructure within these clusters, we also investigated the number and size distribution of subclusters identified as a function of the 
cutoff distance used. For a random distribution, the number of groups identified in the minimum spanning tree is a function of both the minimum group size and the 
cutoff distance. A long cutoff distance and small minimum group size will break the MST into a large number of small, more dispersed clusters. We examined 
the number and size distributions of clusters that would arise in a randomly distributed population of YSOs to find the cutoff distance and minimum group size 
that yielded the largest number of groups that had a low probability of occurring randomly.

We found that using a break length of 40$\arcsec$ (2.2~pc in W49) and requiring $N$~$\geq$~6 or more members isolates seven subclusters, a distribution that did not occur in the 1000 simulated 
randomly distributed trials. Only $8.8\%$ of the random trials had at least one group of six or more members, and only $0.3\%$ of the random YSO distributions had two groups, with no trials 
having more than two random groups. Therefore, the random trials indicate that the identified subclusters likely represent physically associated groups of YSOs, with a $<$10\% likelihood 
that one of the subclusters could be a random association. Note also that the subclusters identified are all within the three clusters identified with the larger break length and 
minimum cluster member number. However, owing to the complex line of sight toward W49, without additional data to determine the distance to each object, we cannot rule out the possibility
that all of the clusters may not all be at the same distance as the W49 GMC.

According to the YSO clustering analysis, identified subclusters in W49 have sizes of order a few parsecs in diameter with six to nine sources in 
each subcluster. On the other hand, subclusters identified in the Cygnus-X DR21 region have 10$-$148 sources with diameters of 1$-$8.1~pc \citep{bee10}. The same study finds subclusters 
in the AFGL 2636 region with diameters of 1.1$-$3.6~pc with 12$-$149 sources. Similarly, \citet{koe08} found subclusters in the W5 region with 10$-$201 sources with diameters of a few parsecs. 
W49 has smaller clusters with less members; however, this might be a result of incompleteness. It also should be noted that these studies used different criteria in their subcluster analyses. To compare  W49  to other previously observed regions, in Section \ref{sec:comp} we examine the clustering properties of the star-forming regions G305 and G333 with the same method 
used for W49 (after rescaling to account for their different distances) and then compare their clustering properties in detail.

\begin{figure*}
\centering
\plotone{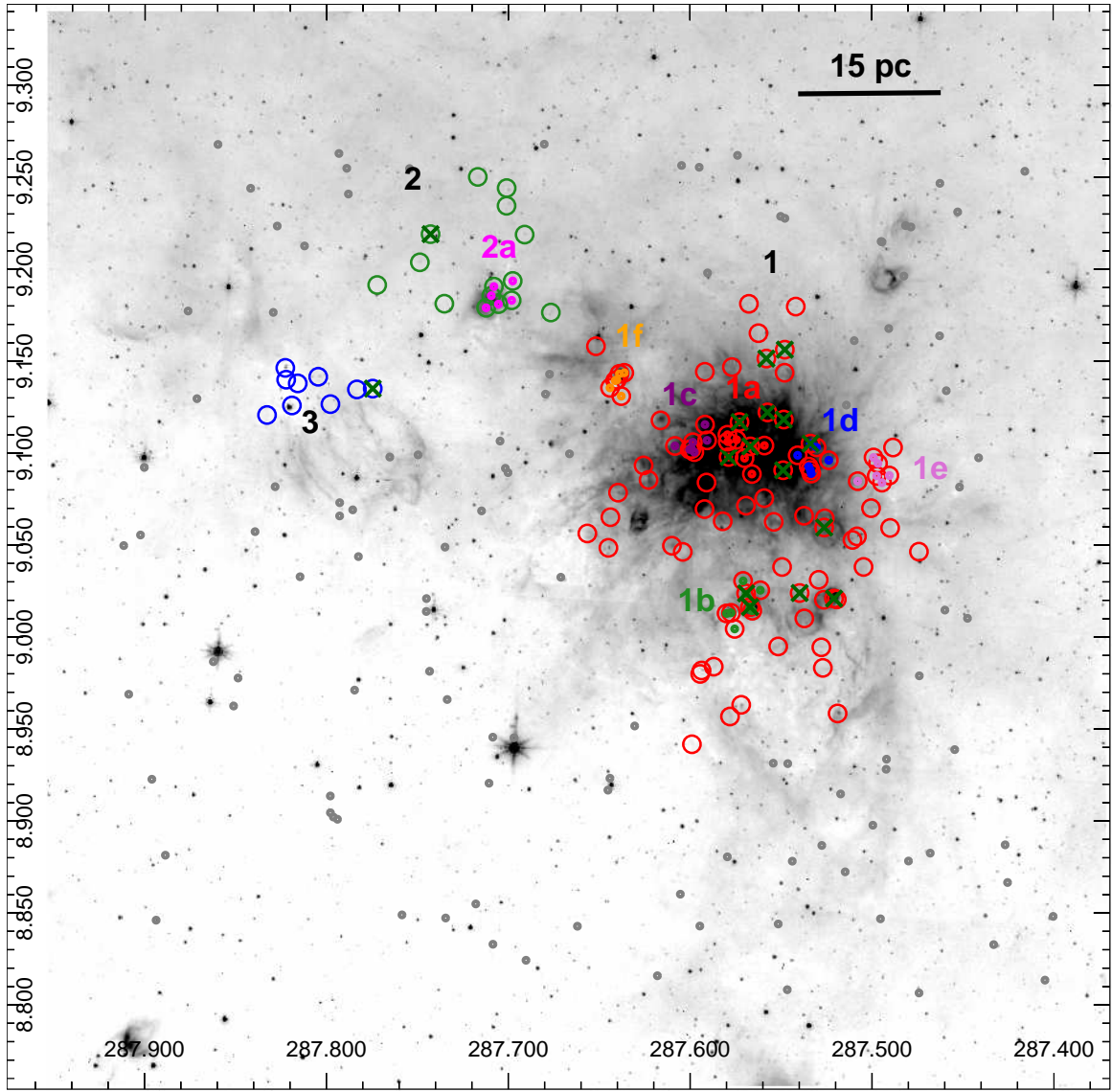}
\caption{Three MST clusters identified using $d_c$ = 96$\arcsec$ are shown with black group numbers overlaid on the IRAC 8~$\mu$m image. The seven subgroups identified by $d_c$ = 40$\arcsec$ are shown with a letter and the cluster number they belong to (see Table~\ref{GMCclusters1} and Table~\ref{GMCclusters2}). 
Massive YSO candidates ($M$ $>$ 8 M$_\odot$) are shown with green crosses. Distributed YSO candidates not assigned to groups are plotted with gray points.}\label{fig:mstresults2}
\end{figure*}

The properties of each W49 subcluster are shown in Table~\ref{GMCclusters2}, and subclusters are shown in the right panel of Figure~\ref{fig:mstresults1}. 
All identified groups and clusters from both large and small cutoff distances can be seen in Figure~\ref{fig:mstresults2} and the summary of the results 
can be seen in Table~\ref{clustersumW49}, with the physical scale assuming that the clusters are at 11.1~kpc.

\begin{deluxetable*}{llcccccccccc}
\tabletypesize{\scriptsize}
\tablecaption{Properties of the Clusters in W49 GMC\label{GMCclusters1} for $d_{c}$ = 96$\arcsec$}
\tablewidth{0pt}
\tablehead{\colhead{No.} & \colhead{Cluster} & \colhead{R.A. (J2000.0)} & \colhead{Decl. (J2000.0)} & \colhead{N$_{IR}$\tablenotemark{a}} & \colhead{I} & \colhead{II} & \colhead{II/I\tablenotemark{b}} & \colhead{N$_{emb}$\tablenotemark{c}} & \colhead{N$_{td}$\tablenotemark{d}} & \colhead{Diameter} & \colhead{Diameter}\\
\colhead{} & \colhead{Name} & \colhead{($\degr$)} & \colhead{($\degr$)} & \colhead{}  & \colhead{}  & \colhead{}  & \colhead{}  & \colhead{} & \colhead{} & \colhead{($\arcsec$)} & \colhead{(pc)}}
\startdata
1	&	G43.15-0.01	&	287.56496	&	9.074502	&	97	& 31	&	57	&	1.84(0.41)	&	9	&	0	&	497.88	&	26.86	\\
2	&	G43.33-0.08	&	287.71438	&	9.202276	&	15	& 2	&	10	&	5.00(3.87)	&	1	&	2	&	244.17	&	13.17	\\
3	&	G43.31-0.20	&	287.808112	&	9.134392	&	9	& 2	&	7	&	3.50(2.81)	&	0	&	0	&	176.38	&	9.52	\\
\enddata
\tablenotetext{a}{Number of stars with infrared excess. Includes Class I, Class II, deeply embedded protostars, and transition disk candidates.} 
\tablenotetext{b}{Number in parentheses indicates Poisson uncertainty in ratio.} 
\tablenotetext{c}{Number of deeply embedded protostars.} 
\tablenotetext{d}{Number of transition disk candidates.}
\end{deluxetable*}

\begin{deluxetable*}{llcccccccccc}
\tabletypesize{\scriptsize}
\tablecaption{Properties of the Clusters in W49 GMC\label{GMCclusters2} for $d_{c}$ = 40$\arcsec$}
\tablewidth{0pt}
\tablehead{\colhead{No.\tablenotemark{a}} & \colhead{Cluster} &  \colhead{R.A. (J2000.0)} & \colhead{Decl. (J2000.0)} & \colhead{N$_{IR}$\tablenotemark{b}} & \colhead{I} & \colhead{II} & \colhead{II/I\tablenotemark{c}} & \colhead{N$_{emb}$\tablenotemark{d}} & \colhead{N$_{td}$\tablenotemark{e}} & \colhead{Diameter} & \colhead{Diameter}\\
\colhead{} & \colhead{Name} & \colhead{($\degr$)} & \colhead{($\degr$)} & \colhead{}  & \colhead{}  & \colhead{}  & \colhead{}  & \colhead{} & \colhead{} & \colhead{($\arcsec$)} & \colhead{(pc)}}
\startdata
1a	&	G43.17-0.00	&	287.571256	&	9.102715	&	9	&	6	&	3	&	0.50(0.35)	&	0	&	0	&	70.53	&	3.81	\\
1b	&	G43.10-0.04	&	287.57085	&	9.017707	&	8	&	3	&	3	&	1.00(0.82)	&	2	&	0	&	88.15	&	4.76	\\
1c	&	G43.19-0.02	&	287.597988	&	9.105411	&	7	&	1	&	6	&	6.00(6.48)	&	0	&	0	&	34.26	&	1.85	\\
1d	&	G43.15+0.03	&	287.532889	&	9.096616	&	7	&	1	&	6	&	6.00(6.48)	&	0	&	0	&	50.86	&	2.74	\\
1e	&	G43.13+0.06	&	287.497549	&	9.089443	&	6	&	1	&	3	&	3.00(3.46)	&	2	&	0	&	60.49	&	3.26	\\
1f	&	G43.23-0.05	&	287.639709	&	9.138874	&	6	&	2	&	4	&	2.00(1.73)	&	0	&	0	&	38.57	&	2.08	\\
2a	&	G43.31-0.08	&	287.705104	&	9.185597	&	6	&	1	&	4	&	4.00(4.47)	&	1	&	0	&	79.40	&	4.28	\\
\enddata
\tablenotetext{a}{Subclusters 1a, 1b, 1c, 1d, 1e, and 1f correspond to cluster 1; subcluster 2a corresponds to cluster 2. Clusters 1, 2, and 3 are determined by the cutoff distance $d_{c}$ = 96$\arcsec$}
\tablenotetext{b}{Number of stars with infrared excess. Includes Class I, Class II, deeply embedded protostars, and transition disk candidates.} 
\tablenotetext{c}{Number in parentheses indicates Poisson uncertainty in ratio.}
\tablenotetext{d}{Number of deeply embedded protostars.} 
\tablenotetext{e}{Number of transition disk candidates.} 
\end{deluxetable*}

\begin{deluxetable}{lll}
\tabletypesize{\scriptsize}
\tablecaption{Clusters/Subclusters in W49 \label{clustersumW49}}
\tablewidth{0pt}
\tablehead{\colhead{Parameter} & \colhead{Straight-line} & \colhead{$d_c$=40$\arcsec$}\\
\colhead{} & \colhead{Fit} & \colhead{}}\\
\startdata
Number of clusters & 3 & 7\\
Cutoff distance & 96$\arcsec$(5.2~pc) & 40$\arcsec$(2.2~pc)\\
Percent in clusters & 52 & 21\\
Group size/diameter & 176$\arcsec$-498$\arcsec$ (9-27~pc) & 34$\arcsec$-88$\arcsec$ (1-5~pc)\\
Class II/I Ratio & 2.11(0.43)\tablenotemark{*} & 2.00(0.64)\tablenotemark{*}\\
\enddata
\tablenotetext{*}{Number in parentheses indicates Poisson uncertainty in ratio.}
\end{deluxetable}

\begin{figure*}[ht!]
\centering
\includegraphics[width=16cm]{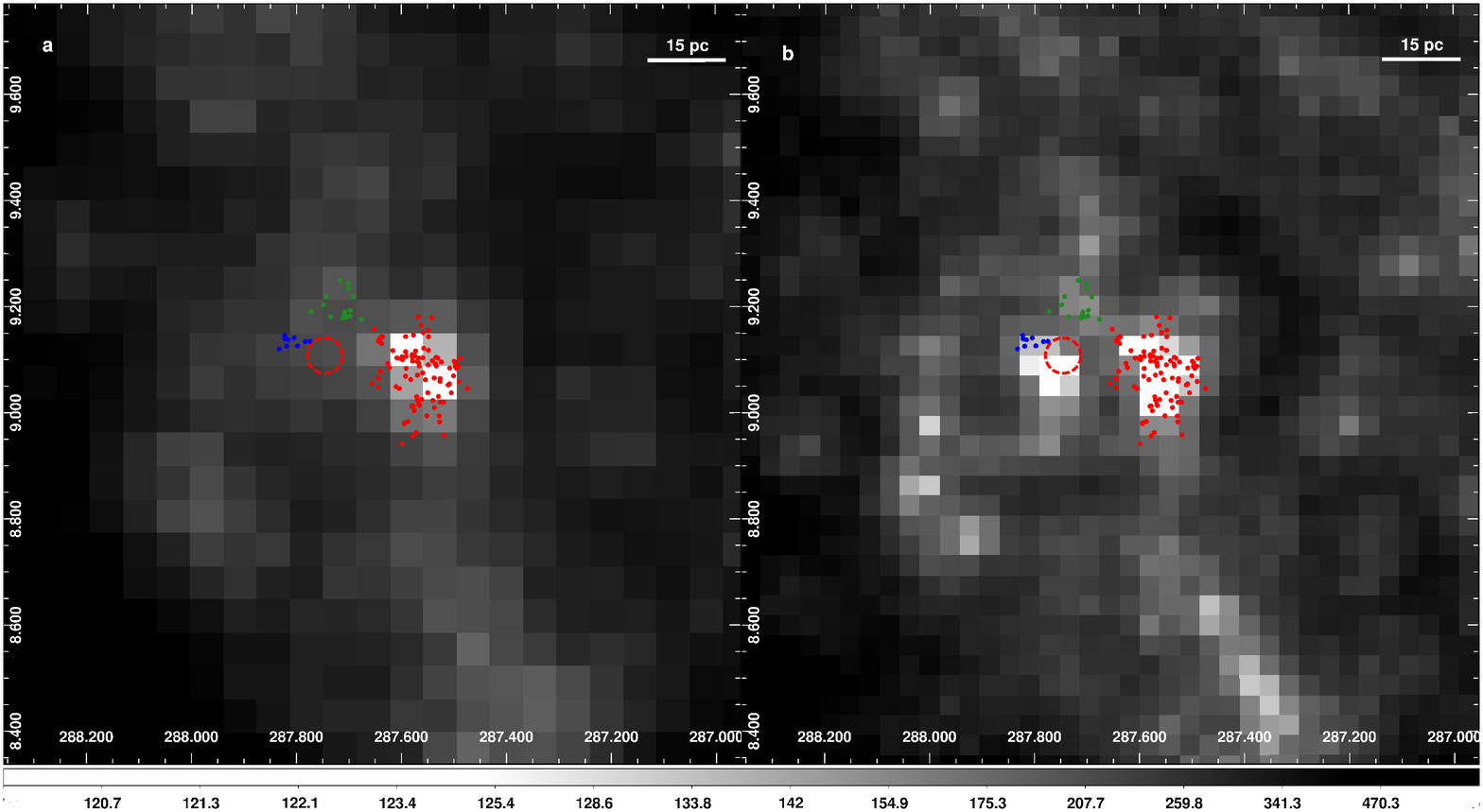}
\caption{YSO nearest-neighbor distance (in arcseconds) in the W49 cloud complex, with the YSO candidates in identified clusters overlayed on each panel. Panel (a) shows the YSO nearest neighbor distance in the W49 cloud complex. 
Panel (b) shows the YSO nearest-neighbor distance including the faint YSOs/possible contaminants. The position of the W49B supernova remnant is indicated by the red circle.}\label{fig:densitymaps}
\end{figure*}

\subsubsection{YSO Densities in Clusters} \label{sec:ysodensity}

Figure~\ref{fig:densitymaps} shows the YSO nearest-neighbor distance in the whole cloud complex, with the YSO candidates in identified clusters overlayed 
on each panel. The YSO nearest-neighbor distance grayscale images show the YSO nearest-neighbor distance in arcseconds, with white regions being higher density. Panel (a) shows the 
plots with the YSO candidate population, and Panel (b) shows the same maps but includes the faint YSOs/possible contaminant sources. It can be seen in Panel (b) 
that there is an overdensity of faint YSOs at the position of W49B, potentially indicating an associated low-mass star cluster, although it could be a chance 
alignment.

\begin{figure*}
\centering
\includegraphics[width=10cm]{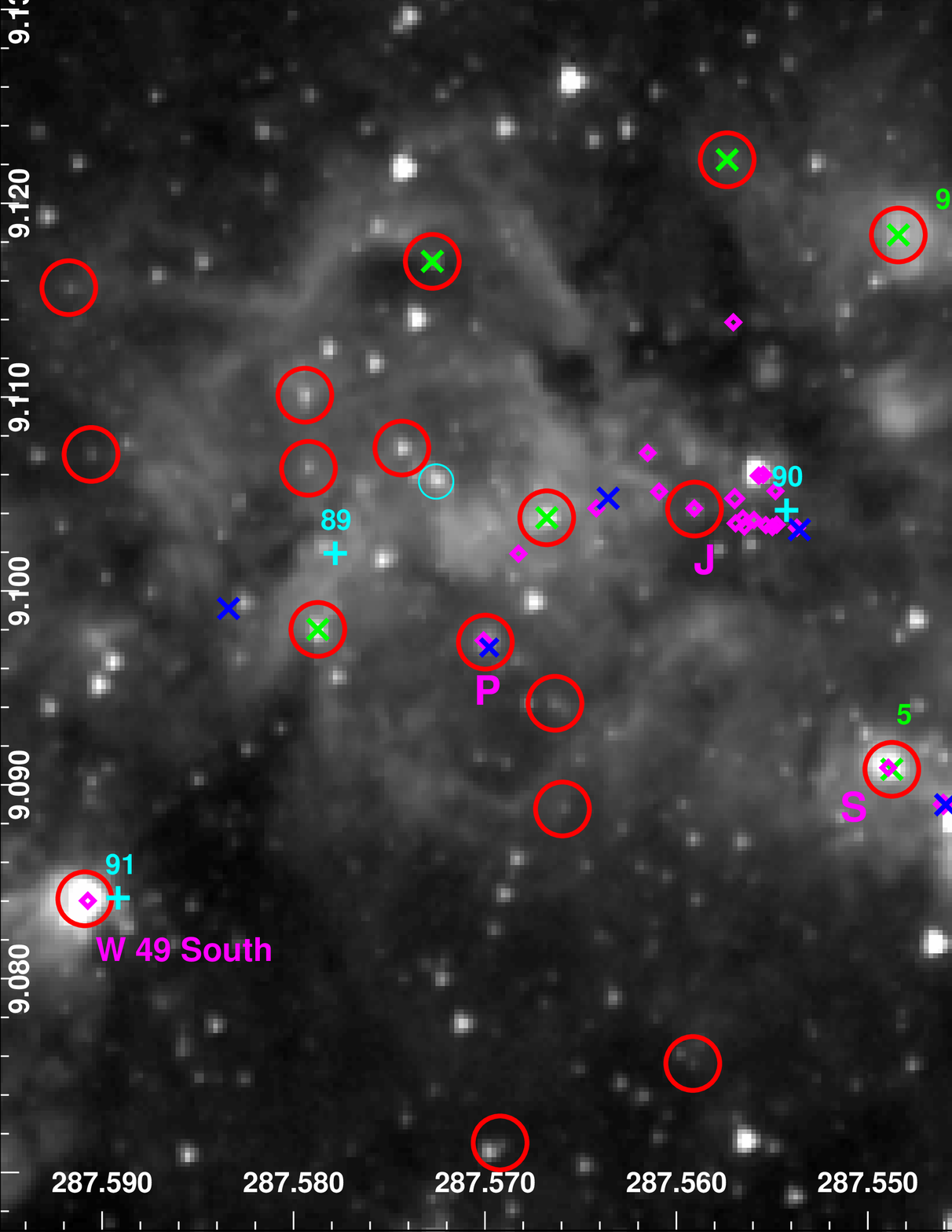}
\caption{Positions of UCHII regions \citep{dep97} in the W49 field are shown with magenta diamonds on the IRAC 3.6~$\mu$m image with the associated YSOs. Red circles show the YSOs in subcluster 1a, and blue circles show the YSOs in subcluster 1d. The cyan circle marks the location of the O type star discovered by \citet{wus14}. Green crosses mark the massive YSO candidates, cyan crosses 89, 90, 91, and 92 mark the massive dust clumps from \citet{mat09}, and blue crosses mark the methanol masers identified by \citet{bre15}.}\label{fig:HIIs}
\end{figure*}

\section{SED Models of MYSOs} \label{sec:sed}

We have applied an SED fitting method described in \citet{azi15} to the 231 YSO candidates in the MST clusters in order to 
identify the most massive candidates in our sample and obtain a rough estimate of their properties. The most common method to find an SED model is to compare 
the models to data points, find the ${\chi}^2$ value using the photometric errors, and pick the model that minimizes the ${\chi}^2$. This best-fit method is 
described in detail by \citet{rob07}.

With a limited amount of photometric data, in order to explore all the possible solutions, we first applied the \citet{azi15} method, which uses a Markov 
Chain Monte Carlo method to explore the larger parameter space of the model grid. We applied this method to the set of YSO candidates and found that 78\% have masses 
$\leq$ 8 M$_\odot$, and 60\% of the YSO candidates have masses between 2 and 6 M$_\odot$.

Within the set of MYSOs found to be $\geq$ 8 M$_\odot$, we selected the ones that have at least six photometric data points to limit the fitting to the 
better-constrained objects, and we used the \citet{rob07} online SED fitter to obtain the physical parameters from the best fit. This selection also favors the 
brighter and possibly more massive objects in our sample. We identified 16 MYSO candidates with this selection method, which are shown in Figure~\ref{fig:mstresults2}
with green crosses.
 
The SED fitter was used to model the available data (2MASS/UKIDSS $J$, $H$, and $K_{s}$, \textit{Spitzer}/IRAC$3.6$, $4.5$, $5.8$, and $8.0$~$\mu$m, and 
\textit{Spitzer}/MIPS $24$~$\mu$m). We focused on only the mass, luminosity, and age parameters; all of the modeled parameters are likely to have a high uncertainty without a finer 
sampling of the object's SED. To provide some indication of the model accuracy, we report the results from the average of the best 10 fit models and their standard deviation for the 
YSO candidates that have masses larger than 8 M$_\odot$ in Table~\ref{SEDresults}. 

Most of the MYSO candidates are classified as Class I, which shows that they are recently formed. MYSO candidates 1 and 16 were also previously identified as YSOs in the Red MSX 
Source Survey \citep[RMS;][]{urq14}.  An \ion{H}{2} region S at the position of MYSO 5 and another \ion{H}{2} region at the position of MYSO 9 were previously identified in the same survey. 
Within 15$\arcsec$ near MYSO 5 there is another \ion{H}{2} region and a methanol maser, which are both indicating that there is ongoing massive star formation (Figure~\ref{fig:HIIs}).

\section{Massive Star Formation Tracers} \label{sec:tracers}

A large fraction of stars and clusters form mostly at the peripheries of \ion{H}{2} regions, and it has been suggested that \ion{H}{2} regions can enhance and 
trigger the massive star formation \citep[e.g.,][]{deh06, deh08,zav06}. Since W49 is a massive star-forming complex that hosts many UCHII regions, 
it is expected to have massive star formation around those locations. As can be seen in Figure~\ref{fig:HIIs}, most of the \ion{H}{2} regions are located within a 1 pc area around 
the Class I candidate SSTOERC G043.1684+00.0087, where the UCHII region \object{W49 J} \citep{dep97} is located. Four of the \ion{H}{2} regions seem positionally associated with YSO candidates, 
including one MYSO candidate. The close correspondence between some of the YSO candidates and the UCHII regions may indicate that some of the YSOs may be related to the driving object behind these UCHII regions,
or at least formed in the same cluster.

Embedded UCHII regions and MYSOs have similar mid-IR colors and very similar ages \citep{mot11}; however, we do not see more associations, contrary to what we expect
from these massive star formation tracers. Several of the \ion{H}{2} regions are very bright in the IRAC images, and a couple are saturated in the $8.0$~$\mu$m band. However, most of the \ion{H}{2} regions do not have an associated YSO
candidate. In some cases we were not able to classify these sources because of the lack of detection in at least four of our photometric bands. For example, it can be seen in Figure~\ref{fig:HIIs} that many of the \ion{H}{2} regions close to W49 J do not have bright mid-IR counterparts, and the sources in this region would be difficult to separate at this resolution. It also can be seen that many of the IR-identified candidate MYSOs do not have associated UCHII or dust clump sources, indicating that the sources are perhaps in different evolutionary stages. 

In addition to HII regions, methanol masers are also known to be associated with high-mass star-forming regions \citep{pes02, bre13}. 
The positions of the 6.7 GHz methanol masers identified by \citet{bre15} in W49 (G43.149+0.013, G43.165+0.013, G43.167-0.004, G43.171+0.005, and G43.175-0.015) are shown in Figure~\ref{fig:HIIs} with blue crosses. These methanol masers appear as unclassified in our catalog according to their lack of photometry; however, G43.167-0.004 seems positionally associated with \ion{H}{2} region P and/or one of our Class I YSO candidates which does not appear to be an MYSO candidate according to the SED modeling. Three of the methanol masers seem also positionally associated with previously identified \ion{H}{2} regions, as can be seen in Figure~\ref{fig:HIIs}. These sources are perhaps in a younger stage of evolution where we cannot yet trace them with IR photometry. We can conclude that high-resolution observations in the infrared, submillimeter, and radio are necessary to put together a comprehensive picture of the star formation in the central region of W49.

\begin{deluxetable*}{lllllllllll}[ht!]
\tabletypesize{\scriptsize}
\tablecaption{Physical Parameters of Massive YSO Candidates\label{SEDresults}}
\tablewidth{0pt}
\tablehead{\colhead{No.} & \colhead{Name} &  \colhead{R.A.} & \colhead{Decl.} & \colhead{Mass} & \colhead{$\sigma$} & \colhead{Luminosity} & \colhead{$\sigma$} & \colhead{Age} & \colhead{$\sigma$} & \colhead{Class}\\
\colhead{} & \colhead{} & \colhead{J2000.0($\degr$)} & \colhead{J2000.0($\degr$)} & \colhead{(M$_\odot$)} & \colhead{(M$_\odot$)} & \colhead{($10^3$L$_\odot$)} & \colhead{($10^3$L$_\odot$)} & \colhead{($10^6$yr)} & \colhead{($10^6$yr)} & \colhead{}}\\
\startdata
1 & SSTOERC G043.0772+00.0038 & 287.520782 & 9.021031 & 12.45 & 1.29 & 12.77 & 3.61 & 0.84 & 0.76 & 1\\
2 & SSTOERC G043.0942$-$00.0388 & 287.566986 & 9.016427 & 10.15 & 1.99 & 3.41 & 1.08 & 0.11 & 0.15 & 1\\
3 & SSTOERC G043.1017$-$00.0373 & 287.569153 & 9.023827 & 9.57 & 0.87 & 5.51& 1.74 & 2.25 & 0.70 & 1\\
4 & SSTOERC G043.1139+00.0169 & 287.526184 & 9.059647 & 7.35 & 0.50 & 0.74 & 0.19 & 0.08 & 0.03 & 2\\
5 & SSTOERC G043.1518+00.0115 & 287.548737 & 9.090796 & 18.96 & 3.61 & 34.37 & 23.66 & 0.02 & 0.02 & 1\\
6 & SSTOERC G043.1578+00.0314 & 287.533691 & 9.105292 & 10.41 & 2.09 & 5.38 & 2.46 & 0.16 & 0.15 & 1\\
7 & SSTOERC G043.1716+00.0017 & 287.566772 & 9.103776 & 8.90 & 0.63 & 4.23 & 0.96 & 1.30 & 0.22 & 1\\
8 & SSTOERC G043.1719$-$00.0114 & 287.578735 & 9.098016 & 11.13 & 2.14 & 9.48 & 6.26 & 0.37 & 0.49 & 1\\
9 & SSTOERC G043.1761+00.0245 & 287.548401 & 9.118363 & 14.14 & 3.26 & 20.10 & 11.01 & 0.48 & 0.31 & 1\\
10 & SSTOERC G043.1837+00.0185 & 287.557343 & 9.122244 & 12.52 & 2.23 & 10.01 & 3.58 & 0.16 & 0.12 & 1\\
11 & SSTOERC G043.1860+00.0026 & 287.572754 & 9.117006 & 13.37 & 2.40 & 16.46 & 7.27 & 0.52 & 0.39 & 1\\
12 & SSTOERC G043.2096+00.0426 & 287.547821 & 9.156357 & 8.09 & 0.74 & 3.27 & 0.84 & 2.53 & 0.94 & 1\\
13 & SSTOERC G043.2099+00.0315 & 287.557983 & 9.151517 & 14.60 & 0.00 & 19.20 & 0.00 & 1.30 & 0.00 & 2\\
14 & SSTOERC G043.2942$-$00.1658 & 287.774597 & 9.135176 & 12.03 & 4.35 & 13.96 & 8.54 & 0.93 & 0.59 & 2\\
15 & SSTOERC G043.3542$-$00.0991 & 287.742798 & 9.219168 & 9.63 & 0.00 & 5.50 & 0.00 & 3.0 & 0.00 & 3\\
16 & SSTOERC G043.0885$-$00.0114 & 287.539703 & 9.024038 & 17.30 & 3.45 & 34.03 & 16.18 & 0.17 & 0.11 & 1\\
\enddata
\end{deluxetable*}

\section{Discussion} \label{sec:results}

\subsection{Star Formation in W49 GMC} \label{sec:history}

\citet{alh03} found four clusters associated with radio sources W49S, W49N, S, and Q and hypothesized that the GMC collapsed to form the central massive 
cluster (W49N) and stellar winds and UV radiation triggered the formation of the Welch ring. However, they did not find any evidence of triggering for other 
clusters on the south and east part of the region (W49S, S, and Q). Also, they indicate that compact \ion{H}{2} regions with short lifetimes can be found over 
the entire region, providing evidence for multiscaled, largely coeval star formation in W49A. In our clustering analysis, we found that the large 
cluster 1 (G43.15-0.01) represents almost the whole GMC, including the well-known Welch ring, W49S, and clusters S and Q.

There is significant correspondence between our YSO subclusters and the clusters identified by \citet{alh03}. Subcluster 1a (G43.17-0.00) corresponds to their 
cluster 1 (``Extended''), while their cluster 2, W49A South, is unresolved in the IRAC images and detected as a single YSO candidate (SSTOERC G043.1651$-$00.0285). 
Their cluster 3, corresponding to UCHII region S, is also identified as a single high-mass Class I YSO candidate in our data, SSTOERC G043.1518+00.0115. 
Their cluster 4, corresponding to UCHII region Q, is bright in the IRAC images but did not meet the YSO selection criteria.

The other subclusters we have identified extend to the east and west of W49A and do not correspond to previously identified near-infrared stellar clusters. 
In total we identified seven subclusters of YSOs within 30~pc of W49A using the MST method. Subclusters 1a, 1b, 1c, 1d, 1e,  and 1f correspond to cluster 1 which has a 
Class II/I ratio of 1.87. Within this cluster, the youngest subcluster 1a, corresponding to the central region of W49A, and subclusters 1c and 1d lie near the outside edge of the 
mid-infrared double-ring structure proposed by \citet{pen10}. The double-ring structure was interpreted as sites of massive star formation triggered by feedback from the central 
stars in W49A. Because of the large uncertainty in the Class II/I ratios due to small number statistics, the relative ages of subclusters according to the ratio of II/I will not be discussed.

Although clusters 2 and 3 do not have significantly more members than clusters that appear in our random trials, we note that both of these clusters contain MYSO candidate objects, which 
do not appear outside of our identified clusters. Cluster 2 contains one of the subclusters (2a) that are significant and do not appear outside of the main clusters found at the larger 
break lengths. It is also associated with a region of enhanced extended IR emission that can be seen in Figure~\ref{fig:GMC}. Cluster 3 is adjacent to and may be associated with the 
supernova remnant W49B. The class II/I ratio of cluster 3 is 3.50$\pm$2.81 and may indicate an apparent age intermediate between cluster 1 and cluster 2.

\citet{mat09} identified dust clumps associated with W49 molecular clouds with continuum emission observations at 850~$\mu$m with 15$\arcsec$ resolution 
by using \textit{SCUBA/James Clerk Maxwell telescope}. The dust clumps 98 and 100 seem associated with the subcluster 1b (G43.10$-$0.04) with their given distances between 11.1 and 11.4~kpc 
determined from $^{13}$CO(1-0) and HI data by \citet{mat09}. The rest of the dust clumps, 83, 84, 85, 86, 87, 89, 90, 91, 92, 93, 94, 95, 96, 97, 101, 102, 
103, and 104, correspond to the region where we identified cluster 1, and they are shown in Figure~\ref{fig:CO} with blue crosses. The dust clumps 89, 90, 91,
and 92, which are shown in Figure~\ref{fig:HIIs}, have the biggest masses between 3.5 $\times10^4$ $-$2.8 $\times10^5$ M$_\odot$, and it was noted by \citet{mat09} that assuming 
a dust temperature of 15 K, these massive clumps will eventually form star clusters via fragmentation. Dust clumps shown with orange crosses in Figure~\ref{fig:CO} were determined 
to be at distances between 4.03 and 7.45~kpc by \citet{mat09}, which might indicate that the clusters 2 and 3 are not associated with the W49 complex. Follow-up observations will 
be necessary to verify the classification and distance of these YSO candidates and confirm their cluster membership.

\subsection{Line-of-sight Cloud and W49 MST Clusters} 

\begin{figure*}[ht!]
\centering
\includegraphics[width=18cm]{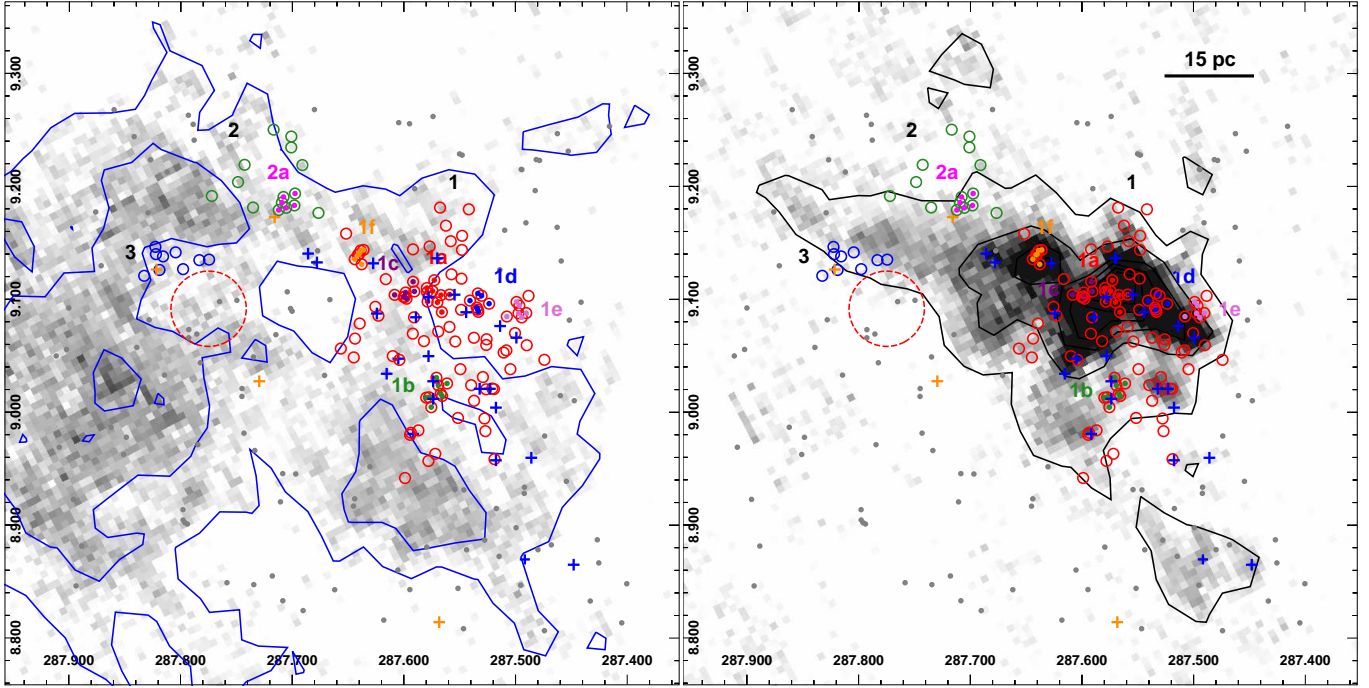}
\caption{$^{13}$CO(1-0) integrated intensity contours \citep{sim01} and YSO clusters overlayed on CO maps in gray scale. Left: $^{13}$CO(1-0) intensity is integrated over the velocity range associated with GRSMC 43.30-0.33 ($\nu_{LSR}$ = 35 $-$ 50 km s$^{-1}$) and contour levels are 2.5, 6.25, 10 K km s$^{-1}$, shown in blue; right: $^{13}$CO(1-0) intensity is integrated over the velocity range associated with W49 GMC ($\nu_{LSR}$ = -5 $-$ 25 km s$^{-1}$), and contour levels are 2.5, 14, 25.5, 37, 48.5 and 60 K km s$^{-1}$ for W49 GMC, shown in black. 
Blue crosses show the dust clumps determined at 11.1$-$11.4~kpc, and orange crosses show the dust clumps determined at 4.10$-$7.30~kpc by \citet{mat09}. The red circle indicates the position of the supernova remnant W49B.}\label{fig:CO}
\end{figure*}

\citet{sim01} showed that the GRSMC 43.30-0.33 cloud, which was identified with the Galactic Ring Survey \citep[GRS;][]{jac06}, is near the line of sight to 
W49 at a distance of 3~kpc and largely overlaps W49. We used the integrated intensity CO maps from \citet{sim01} to compare the spatial distributions of 
YSOs in each cloud. We show the YSO clusters with the CO map contours in Figure~\ref{fig:CO}. Cluster 1 (G43.15-0.01) and the subclusters 1a, 1b, 1c, 1d, 1e, and 
1f seem strongly associated with the strongest CO emission in the W49 GMC. Cluster 2 (G43.33-0.08) does not seem strongly associated with the CO emission, 
and it is roughly in line with a dust clump at a distance of 7.45$-$7.30~kpc \citep{mat09}. However, it contains an MYSO candidate 
(SSTOERC G043.3542$-$00.0991) and subcluster 2a (G43.31$-$0.08) which seems more strongly associated with the CO emission than the rest of the cluster.

Similarly, on the line of sight of cluster 3 (G43.31$-$0.20), there is a dust clump at a distance of 4.37$-$4.10~kpc \citep{mat09} that might be associated with 
the foreground cloud GRSMC 43.30-0.33. The Class I YSO candidate closest to the \citet{mat09} dust clump is offset by 8$\arcsec$ from a group of infrared 
sources that are partially resolved into four peaks of emission within a 4.5$\arcsec$ aperture in the 3.6 and 4.5~$\mu$m images but unresolved at 5.8 and 
8.0~$\mu$m. However, there is an MYSO candidate in that cluster (SSTOERC G043.2942$-$00.1658) that could be associated with the CO emission coming from the W49 cloud, as can be seen in Figure~\ref{fig:mstresults2} and Figure~\ref{fig:CO}. There are no clusters that seem to be associated with the brightest CO 
emission from GRSMC 43.30-0.33 like we see in W49, and with it being closer to us, we would expect to be able to detect lower-mass and lower-luminosity objects from 
clusters associated with the foreground cloud. Because of its projected position on the sky, cluster 3 might be thought to have the greatest chance of being a 
foreground object, potentially associated with GRSMC 43.30-0.33 at 3~kpc, the dust clump at 4.37$-$4.10~kpc, or the W49B supernova remnant region at more 
than 10~kpc, as previously mentioned in Section \ref{sec:ysodensity}.

\subsection{Comparison with Other Star-forming Regions} \label{sec:comp}

The 11.1~kpc distance to W49 limits our sensitivity to detect only the brightest and most luminous YSO candidates. To better understand W49 in the context of other regions forming
rich clusters, we investigate how the YSO populations of luminous star-forming complexes G305 and G333 would appear if they were at the greater distance of W49. 
Taken together, these three regions also probe different evolutionary stages, from deeply embedded sources forming protoclusters to already-revealed open clusters.
  
G305 \citep[$l,b = 305\fdg4, +0\fdg1$;][]{cla04} is one of the most massive \citep[$M$ $\gtrsim$ 6 $\times10^5$M$_\odot$;][]{hin10} star-forming complexes in the Galaxy, and is located 
in the Scutum-Crux arm at an estimated distance of $\sim$4~kpc. The complex has a ringlike morphology of bright mid-infrared emission extending over $\sim$34 pc that surrounds two 
optically visible open clusters, Danks 1 and Danks 2. The G305 complex has a considerable amount of ongoing star formation around the periphery of its oldest object, Danks 2 
($\displaystyle{3^{+3}_{-1}}$ Myr), and the younger Danks 1 ($\displaystyle{1.5^{+1.5}_{-0.5}}$) \citep{dav12}.

G333 (also known as RCW 106; $l, b = 333\fdg3, -0\fdg4$) lies at a distance of 3.6~kpc \citep{gar14} and extends over 50 pc. It contains a number of \ion{H}{2} regions and MYSO candidates, with an estimated bolometric luminosity $\log(L_{bol})=6.28$, making it one of the most luminous massive star-forming regions in the Galaxy \citep{urq14}.

Using the \textit{Spitzer} mid-infrared photometry catalog from \citet{wil15}, we identified 1189 and 1057 YSO candidates in G305 and G333, respectively, by using the classification method 
described in Section \ref{sec:yso}, and performed the clustering analysis as described in Section \ref{sec:mstmethod}. We investigate the clustering properties of G305 and G333 
both at their assumed distances and as they would appear if projected to the same distance as W49.

\begin{figure*}
\centering
\includegraphics[width=18cm]{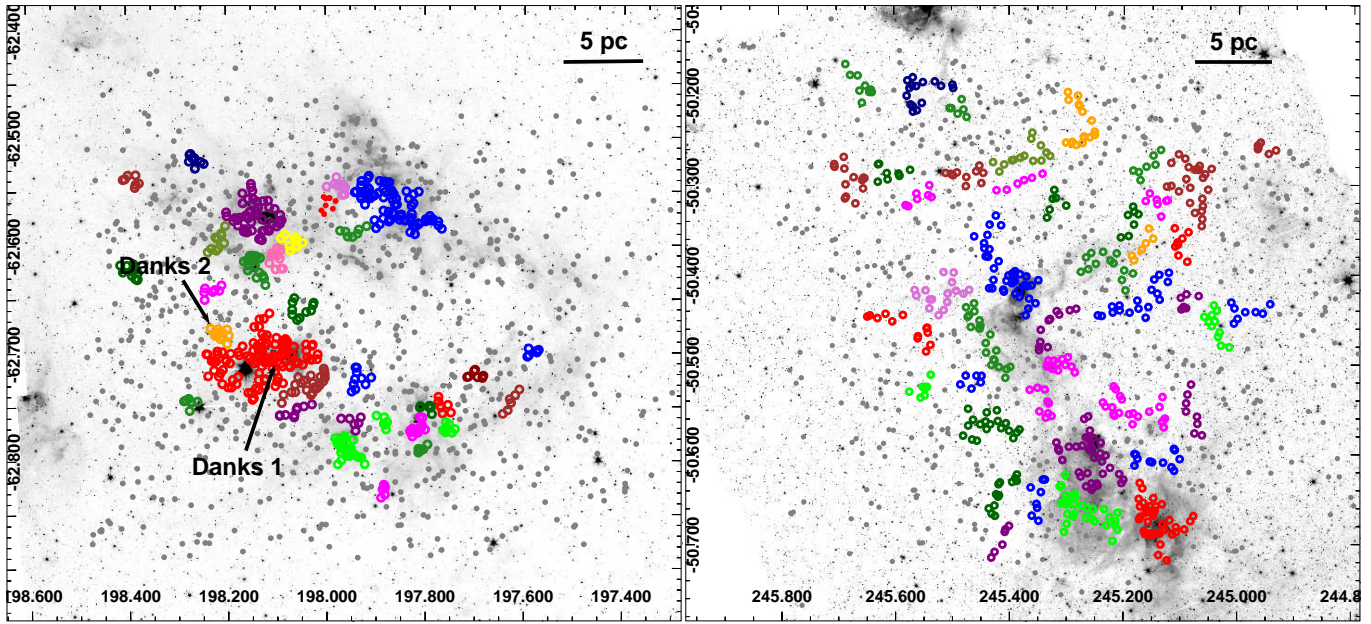}
\caption{Left: subclusters identified in the G305 star-forming region (using $d_c$ =  30$\arcsec$) are overlaid on the IRAC 5.8~$\mu$m image. Right: subclusters identified in the G333 star-forming region (using $d_c$ =  52$\arcsec$) are overlaid on the IRAC 5.8~$\mu$m image. Distributed YSO candidates not assigned to groups are plotted with gray points.}\label{fig:G305andG333clusters}
\end{figure*}

These regions have a rich population of YSOs, and we identified 15 YSO clusters with this MST method in the G305 complex using the straight-line fit method determining 
$d_{c}$ $=$ 48$\arcsec$ and we set $N$ $\geq$ 7 members. With the same method, we found 11 YSO clusters with $d_{c}$ $=$ 73$\arcsec$ in G333. In order to see the hierarchical structure, 
we also used a short cutoff distance that maximizes the number of clusters, identifying the clusters shown in Figure~\ref{fig:G305andG333clusters}. The clustering results for these 
two regions at their original distances are shown in Table~\ref{clustersumG305} and in Table~\ref{clustersumG333}. 

\begin{deluxetable}{lll}
\tabletypesize{\scriptsize}
\tablecaption{Clusters/Subclusters in G305 \label{clustersumG305}}
\tablewidth{0pt}
\tablehead{\colhead{Parameter} & \colhead{Straight-line} & \colhead{$N_{grp}$}\\
\colhead{} & \colhead{Fit} & \colhead{Max1}}\\
\startdata
Number of clusters & 15 & 32\\
Cutoff distance & 48$\arcsec$ (0.9~pc) & 30$\arcsec$ (0.6~pc)\\
Percent in clusters & 76 & 44\\
Group size & 36$\arcsec$-911$\arcsec$ (1-18~pc) & 9$\arcsec$-295$\arcsec$ (0.2-6~pc)\\
Class II/I ratio & 3.5(0.3)\tablenotemark{*} & 3.0(0.3)\tablenotemark{*}\\
\enddata
\tablenotetext{*}{Number in parentheses indicates Poisson uncertainty in ratio.}
\end{deluxetable}

\begin{deluxetable}{lll}
\tabletypesize{\scriptsize}
\tablecaption{Clusters/Subclusters in G333 \label{clustersumG333}}
\tablewidth{0pt}
\tablehead{\colhead{Parameter} & \colhead{Straight-line} & \colhead{$N_{grp}$}\\
\colhead{} & \colhead{Fit} & \colhead{Max1}}\\
\startdata
Number of clusters & 11 & 45\\
Cutoff distance & 73$\arcsec$ (1.3~pc) & 52$\arcsec$ (0.9~pc)\\
Percent in clusters & 81 & 54\\
Group size & 110$\arcsec$-1353$\arcsec$ (2-24~pc) & 43$\arcsec$-306$\arcsec$ (1-5~pc) \\
Class II/I ratio & 4.5(0.4)\tablenotemark{*} & 4.0(0.4)\tablenotemark{*} \\
\enddata
\tablenotetext{*}{Number in parentheses indicates Poisson uncertainty in ratio.}
\end{deluxetable}

For a large branch length cutoff, the probability of the observed clustering properties occurring randomly is relatively high. In G305 the most significant result is the size of the two 
largest clusters, which occured in less than $0.1\%$ of our simulations. In G333 the largest cluster is smaller (378 members) and therefore has a higher probability of 
occurring randomly ($7\%$).

The results for both clusters are more significant with a shorter cutoff distance. The probability of the G333 subclusters occurring randomly is $3.3\%$$-$$18.3\%$, indicating that a 
larger number of these structures may be randomly connected sets of objects. This is supported by the morphology of the G333 subclusters, which display a significant number of 
subclusters with very linear or filamentary appearance that occur randomly. The probability of the number and sizes of G305 subclusters occurring randomly is $<0.1\%$, and the clusters 
also have a strong association with features in the extended IRAC 5.8~$\mu$m image, including subclusters associated with both the open clusters Danks 1 and Danks 2. 

We determined the projected appearance of the G305 and G333 YSOs by rescaling the relative source positions and magnitudes for a distance shift from 3.6~kpc (G333) and 4~kpc 
(G305) to the 11.1~kpc distance to W49. We applied a limiting magnitude of 20.3, 19.3, 18.2, 18.6, 16.3, 14.8, 14.3, and 9.4~mag for $J$, $H$, $K$, $3.6$, $4.5$, $5.8$, $8.0$, and $24$~$\mu$m, respectively, 
which are the faintest detections in the W49 field with error smaller than 0.2~mag. After reprojecting the source coordinates, we did not find any overlapping sources within the same 
photometric aperture (2$\arcsec$) at the rescaled distance.

After rescaling, we reclassified all the sources as described in Section \ref{sec:yso}. We do not assume any additional reddening, so the colors of the objects do not change after 
projecting. However, sources may be below the applied limiting magnitude from the W49 catalog in one band and above in others, resulting in a change in the source classification based on 
the new, more limited wavelength coverage.

The new projected catalog for G305 contains 696 YSOs, only $59\%$ of YSOs from the original catalog, and the projected catalog for G333 contains 404 YSOs, only $38\%$ of the 
original YSOs. In Section \ref{sec:yso} we applied a cut for sources with $[3.6]>13$ to remove likely background objects from the YSO sample. When we apply the same cut to the projected 
G305 YSOs, the resulting population is $6.9\%$ of the original size, and for G333 the resulting YSO population after applying the faint-object cut is $7.9\%$.

The field of view for G305 and G333 was approximately the same at their original distances, 0.5$\degr$~$\times$~0.5$\degr$. After reprojecting G305, the area decreases 
to 0.18$\degr$~$\times$~0.18$\degr$ and the G333 area decreases to 0.16$\degr$~$\times$~0.16$\degr$. We applied the MST clustering analysis to the G333 and G305 YSO populations at both 
their original and projected distances, comparing to simulated random clusters for each set of YSO population and observed or projected area parameters.

In Section \ref{sec:mstw49}, the straight-line fit method for W49 appeared to identify GMC-scale clusters. For the rescaled clusters without cutting off sources that fall below the
$[3.6]=13$~mag cut, we used only the straight-line fit method. For the rescaled G305 we derived the cutoff distance as 22$\arcsec$ (corresponds to $d_c$ = 1.2~pc, at a distance of 
11.1~kpc) and for the rescaled G333 we derived the cutoff distance as 33$\arcsec$ (corresponds to $d_c$ = 1.8~pc, at a distance of 11.1~kpc). 
For G305 and G333 the reprojected straight-line fit cutoff distance finds only one cluster per region, with a clustered fraction of $74\%$ in rescaled G305 and $81\%$ in G333.

Finally, we examined the clusters in the rescaled G305 and G333 as they would appear if we removed the sources fainter than $[3.6]=13$. The straight line-fit cutoff distance 
without faint sources was 52$\arcsec$ for G305 and 78$\arcsec$ for G333. Although these are shorter cutoff distances than W49, the clusters found are not highly significant. 
The largest cluster in rescaled G333 has only 51 members, only half the size of cluster 1 in W49. The chance of the largest rescaled G333 cluster occurring randomly was $3.2\%$, 
compared to $<0.1\%$ for cluster 1 in W49. The G305 rescaled clusters were smaller, but because of their shorter cutoff distance, they were also more significant. The largest cluster in 
rescaled G305 had only 32 members, a size that was met in only $0.2\%$ of our simulated rescaled G305 clusters.

\begin{figure*}
\centering
\includegraphics[width=7.5cm]{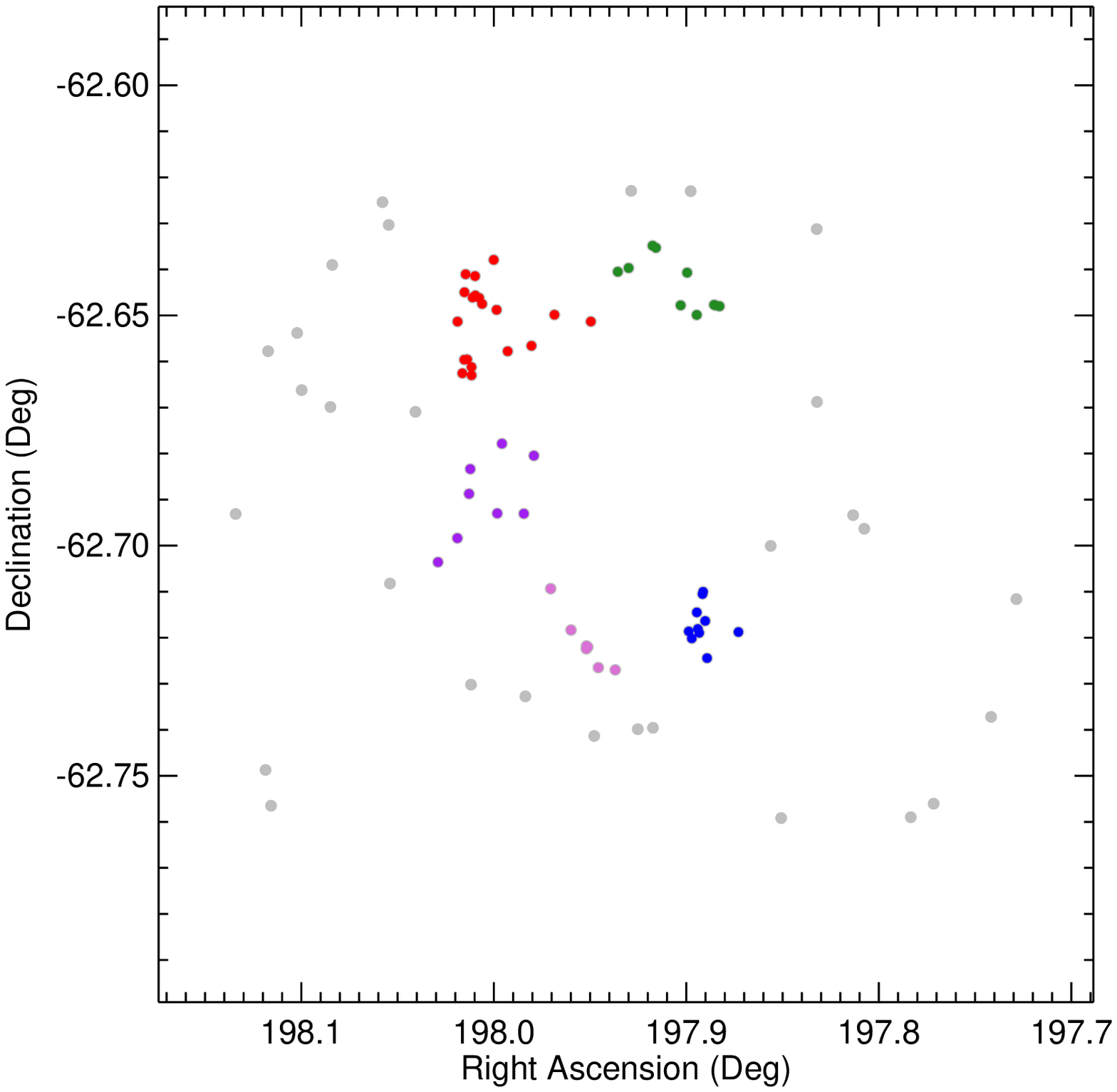}
\includegraphics[width=7.5cm]{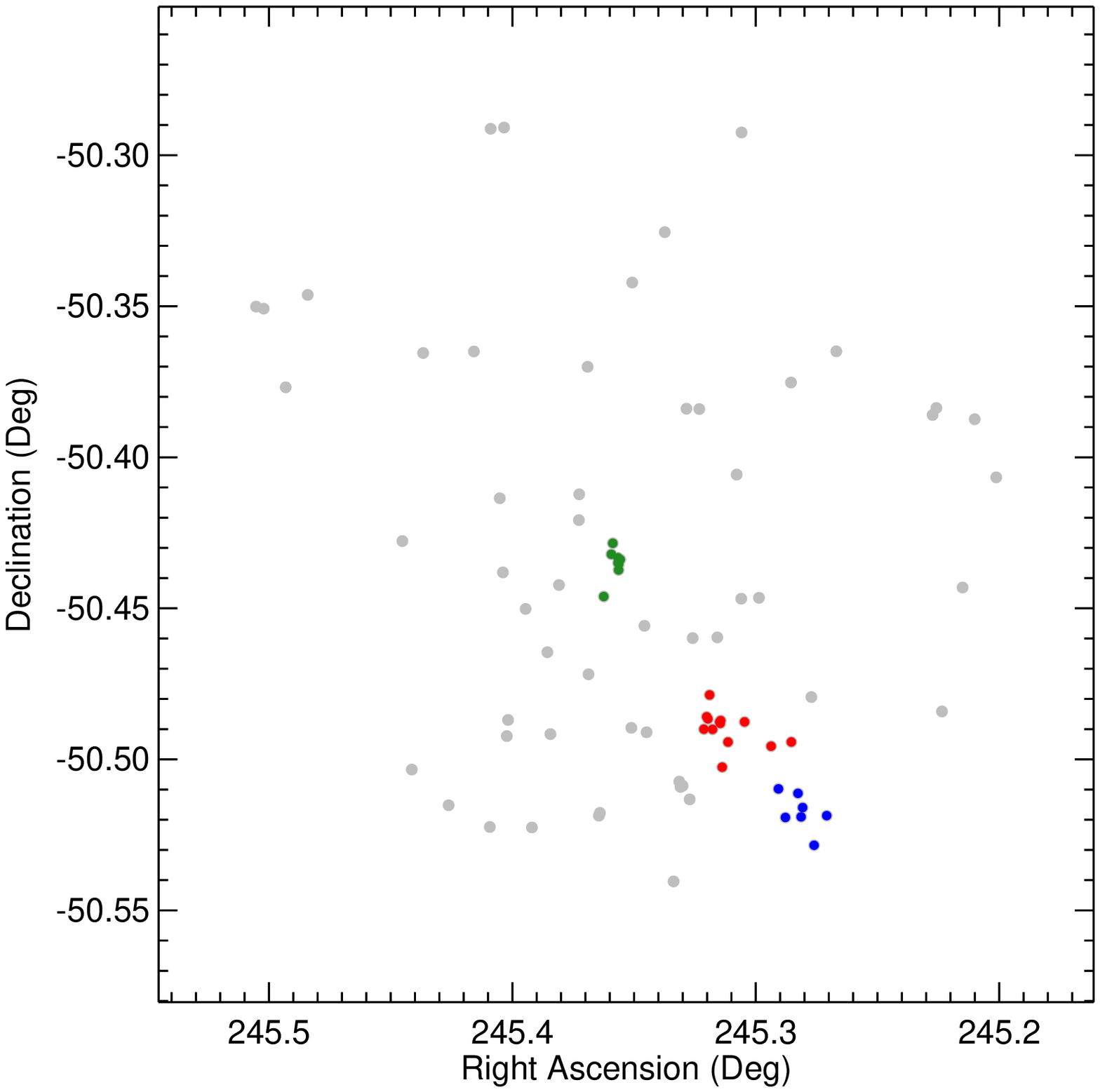}
\caption{Left: subclusters identified in the rescaled G305 star-forming region using the W49 cutoff distance $d_c$ = 40$\arcsec$. The YSOs are plotted in colors according to the clusters identified. The distributed YSO candidates not assigned to groups are plotted with gray points. Right: same as the left panel, but for G333.}\label{fig:rescaledgroups}
\end{figure*}

To directly compare the sizes and numbers of the clusters in W49 to G305 and G333, we also determined the clustering properties in G305 and G333 using the W49 parameters from 
the straight-line fit ($d_c=96\arcsec$ with $N>7$) and the peak that maximized the number of groups ($d_c=40\arcsec$ with $N>6$). We found only one cluster with $d_c=96\arcsec$, and 
although the clusters are larger than for the best-fit cutoff distance for G305 and G333 individually, they are still smaller than the largest cluster in W49. This indicates that W49 
has a larger population of high-luminosity YSOs than either G333 or G305. With $d_c=40\arcsec$ we found three clusters in G333 and five clusters in G305, with similar size 
distributions to W49 clusters 1a$-$1f and 2a, as can be seen in Figure~\ref{fig:rescaledgroups}. There were three clusters in only $0.3\%$ of the rescaled G333 simulations, and there were five clusters in only $0.2\%$ of the rescaled G305 
simulations. The number and size distributions of the subclusters in G305, G333, and W49 are all significant, although W49 again has the richest population, with seven subclusters found.
In W49, six of the seven clusters are found within an area of approximately 0.15$\degr$~$\times$~0.15$\degr$, comparable to the projected area of G305 and G333 and indicating that the larger 
number of tightly packed clusters in W49 is not simply due to the larger field. The W49, G305, and G333 clustering properties and associated probabilities are all summarized in Table~\ref{clustercomp}.

\begin{deluxetable*}{llllll}
\tabletypesize{\scriptsize}
\tablecaption{W49, G305, and G333 Clustering Comparison \label{clustercomp}}
\tablewidth{0pt}
\tablehead{\colhead{Region\tablenotemark{a}} & \colhead{Cutoff Distance} & \colhead{First Cluster Size\tablenotemark{b}} & \colhead{Second Cluster Size} & \colhead{Third Cluster Size} & \colhead{Groups $>$ $N_{min}$\tablenotemark{c}}}\\
\startdata
W49			& 96$\arcsec$	& 97 	& 15 	& 9 	& 3 \\ 
Straight-Line Fit	&		& $<0.1\%$ & $66.6\%$ & $99.3\%$ & $99.9\%$ \\
W49			& 40$\arcsec$	& 9 	& 8 	& 7 	& 7 \\ 
Peak	&		& $0.2\%$ & $<0.1\%$ & $<0.1\%$ & $<0.1\%$ \\
\hline \\
G333  & 73$\arcsec$	 & 378	 & 309	 & 39	 & 11 \\
Straight-line fit 	 & 	 & $7\%$	 & $<0.1\%$	 & $99\%$	 & $99.9\%$ \\
\noalign{\smallskip}
G333 & 52$\arcsec$	 & 40	 & 39	 & 37	 & 45 \\
Peak 	 & 	 & $3.3\%$	 & $<0.1\%$	 & $<0.1\%$	 & $<0.1\%$ \\
\noalign{\smallskip}
G305 	 & 48$\arcsec$	 & 418	 & 328	 & 26	 & 15 \\
Straight-line fit 	 & 	 & $<0.1\%$	 & $<0.1\%$	 & $99.9\%$	 & $99.9\%$ \\
\noalign{\smallskip}
G305 & 30$\arcsec$	 & 113	 & 78	 & 49	 & 32 \\
Peak 	 & 	 & $<0.1\%$	 & $<0.1\%$	 & $<0.1\%$	 & $<0.1\%$ \\
\noalign{\smallskip}
G333 rescaled 	 & 78$\arcsec$	 & 51	 & 6	 & \nodata	 & 2 \\
Straight-line fit	 & 	 & $3.2\%$	 & $99\%$	 & \nodata	 & $99.9\%$ \\
\noalign{\smallskip}
G333 rescaled	 & 52$\arcsec$	 & 24	 & 18	 & \nodata	 & 2 \\
Peak	 & 	 & $<0.1\%$	 & $<0.1\%$	 & \nodata	 & $65\%$ \\
\noalign{\smallskip}
G305 rescaled	 & 52$\arcsec$	 & 32	 & 10	 & 10	 & 4 \\
Straight-line fit	 & 	 & $0.2\%$	 & $29.3\%$	 & $4.9\%$	 & $19.3\%$ \\
\noalign{\smallskip}
G305 rescaled	 & 40$\arcsec$	 & 19	 & 10	 & 10	 & 5 \\
Peak	 & 	 & $0.1\%$	 & $0.2\%$	 & $<0.1\%$	 & $0.2\%$ \\
\noalign{\smallskip}
G333 rescaled 	 & 96$\arcsec$	 & 57	 & \nodata	 & \nodata	 & 1 \\
W49 straight-line dist.	 & 	 & $55.8\%$	 & \nodata	 & \nodata	 & $99.9\%$ \\
\noalign{\smallskip}
G333 rescaled 	 & 40$\arcsec$	 & 13	 & 8	 & 7	 & 3 \\
W49 peak dist.	 & 	 & $0.2\%$	 & $<0.1\%$	 & $0.1\%$ & $0.3\%$ \\
\noalign{\smallskip}
G305 rescaled	 & 96$\arcsec$	 & 74	 & \nodata	 & \nodata	 & 1 \\
W49 straight-line dist.	 & 	 & $85\%$	 & \nodata	 & \nodata	 & $99.9\%$ \\
\noalign{\smallskip}
G305 rescaled	 & 40$\arcsec$	 & 19	 & 10	 & 10	 & 5 \\
W49 peak dist.	 & 	 & $0.1\%$	 & $0.2\%$	 & $<0.1\%$	 & $0.2\%$ \\
\enddata
\tablenotetext{a}{Rescaled G305 and G333 do not contain sources below the $[3.6]=13$ cut}
\tablenotetext{b}{Percentage is showing the probabilty of the clusters to occur randomly.}
\tablenotetext{c}{$N_{min}$=6 for peak and $N_{min}$=7 for straight-line fit}
\end{deluxetable*}

\subsection{Class II/I Ratio as a Cluster Age Indicator}\label{ratiosec}

As seen in Table~\ref{clustersum}, the Class II/I ratios in G305 and G333 change when the YSOs are reprojected to W49 distance, and because of its higher Class II/I 
ratio, W49 appears oldest and the G305 region appears to be the youngest. However, as we mentioned in Section \ref{sec:comp}, G305 has an older population with its two open clusters.
This drop in Class II/I ratio in G305 and G333 regions after rescaling to the greater distance is mainly because lower-luminosity Class II YSOs drop below our detection limits before Class I YSOs do, 
biasing our observations of distant regions toward younger objects. This effect must be taken into account when comparing clouds at different distances using the Class II/I ratio as an age indicator.

\begin{deluxetable}{lccc}
\tabletypesize{\scriptsize}
\tablecaption{Clustering Analysis Summary \label{clustersum}}
\tablewidth{0pt}
\tablehead{\colhead{Region\tablenotemark{a}} & \colhead{Cutoff} & \colhead{Cutoff} & \colhead{Class II/I}\\
\colhead{} & \colhead{Distance($\arcsec$)} & \colhead{Distance (pc)} & \colhead{Ratio\tablenotemark{b}}}\\
\startdata
W49 & 96 & 5.2 & 1.8(0.4)\\
G305 & 48 & 0.9 & 3.5(0.3)\\
Rescaled G305 & 96 & 5.2 & 0.5(0.1)\\
G333 & 73 & 1.3 & 4.5(0.4)\\
Rescaled G333 & 96 & 5.2 & 1.0(0.3)\\    
\enddata
\tablenotetext{a}{W49 parameters from straight-line fit $d_c$ =  96$\arcsec$ have been used here to directly compare the biggest clusters in the rescaled regions without sources fainter than $[3.6]=13$}
\tablenotetext{b}{Number in parentheses indicates Poisson uncertainty in ratio.}
\end{deluxetable}

\section{Summary} \label{sec:sum}

In this paper, we combined \textit{Spitzer} IRAC (3$-$8 $\micron$), MIPS (24 $\micron$), and UKIDSS near-IR ($JHK_S$) data for the W49 GMC to identify and classify 
YSOs by using their infrared colors and magnitudes and analyzed their clustering properties according to their spatial distribution across the region. 
We found the following numbers of YSO candidates: 186 Class I, 907 Class II, 74 transition disks, and 46 deeply embedded protostellar sources. We used the 
MST method to identify the groups and subclusters in the region and found that $52\%$ of YSOs (including transition disk objects) belong to clusters of 
$\geq$7 members in the W49 GMC. In order to assess the significance of the identified MST clusters, we performed simulations on randomly distributed YSOs and examined the 
probability of finding random clusters. We found that cluster 1 represents a large-scale structure in the cloud extending $\sim$27~pc in diameter, with a very low probability 
of occurring in a random distribution of sources. This cluster is centered on the previously identified main W49 region with many tracers of high-mass star formation detected. 
We also used a smaller cutoff distance to investigate the hierarchical structure in the cloud, finding several subclusters within the larger clusters. The distribution of identified 
subclusters has a very low probability of occurring randomly.

We applied an SED fitting method to 231 YSO candidates in the MST clusters in order to identify the most massive candidates in our sample and found 16 YSO candidates 
that have masses $\geq$8 M$_\odot$. We found that the massive Class I candidate SSTOERC G043.1518+00.0115 is located at the position of the UCHII region W49 S. However, we could not find 
any more associations because most of the \ion{H}{2} regions are unclassified in our catalog owing to lack of photometry. We also examined previously identified methanol 
masers in the region that are known as good indicators of high-mass star formation. However, we do not find any IR-identified MYSO candidates that appear physically associated with those masers.

We compared the W49 region to two other star-forming regions, G305 and G333. In our cluster analysis we used the cutoff distances and the ratio of class II/I sources in each cloud in order to derive 
information about their relative ages and YSO densities. In order to directly compare them, we determined the projected appearance of the G305 and G333 regions and compared the 
number and size distributions of the subclusters. We saw that there are more clusters in the W49 region, which indicates that W49 has the richest population of high-luminosity YSOs. However, 
we could not use the ratio of Class II/I objects as an age indicator of clusters, since at greater distances we lose a significant number of older objects, which biases our observations. This comparison also indicates that there is potentially a large number of YSOs in W49 that we have not detected.

\acknowledgments
This work is based in part on observations made with the {\it Spitzer Space Telescope}, which is operated by the Jet Propulsion Laboratory, California 
Institute of Technology, under a contract with NASA. Support for this work was provided by NASA. This publication makes use of data products from the Two 
Micron All Sky Survey, which is a joint project of the University of Massachusetts and the Infrared Processing and Analysis Center/California Institute of 
Technology, funded by the National Aeronautics and Space Administration and the National Science Foundation. The authors gratefully acknowledge the referee for useful comments and 
additional discussion. G. S. acknowledges partial support from NASA Grant NNX12AI60G and Istanbul University grant BAP50195 since this work is part of her PhD thesis research.

{\it Facilities:} \facility{2MASS ($JHK_S$)}, \facility{Spitzer} (IRAC, MIPS), \facility{UKIRT ($JHK_S$)}

\end{document}